\documentstyle[11pt]{article}
\newcommand{\vsig}{\mbox{\boldmath $\sigma$ \unboldmath}}
\newcommand{\veps}{\mbox{\boldmath $\epsilon$ \unboldmath}}
\newcommand{\valf}{\mbox{\boldmath $\alpha$ \unboldmath}}
\newcommand{\vtau}{\mbox{\boldmath $\tau$ \unboldmath}}  
\newcommand{\vpi}{\mbox{\boldmath $\pi$ \unboldmath}}  

\begin{document}
\bibliographystyle{unsrt}

\title{\bf Vector Meson Photoproduction with an Effective Lagrangian
 in the Quark Model }

\author{Qiang Zhao, Zhenping Li \\  
Department of Physics, Peking University, Beijing, 100871, P.R.China\\
C. Bennhold\\
Department of Physics, Center for Nuclear Studies,\\
 The George Washington University, Washington, D.C., 20052, USA}  
\maketitle  
  
\begin{abstract}
A quark model approach to the photoproduction of vector mesons 
off nucleons is proposed.  Its starting point is an effective 
Lagrangian of the interaction between the vector meson and 
the quarks inside the baryon, which generates the non-diffractive 
s- and u- channel resonance contributions. Additional t-channel $\pi^0$ 
and $\sigma$ exchanges are included for the $\omega$ and $\rho^0$
production respectively to account for the large diffractive behavior in
the small $t$ region as suggested by Friman and Soyeur. The numerical results
are presented for the $\omega$ and $\rho$ productions in four
isospin channels with the same set of parameters, and they are in good
agreement with the available data not only in $\omega$ and $\rho^0$
productions but also in the charged $\rho$ productions where the additional
t-channel $\sigma$ exchange does not contribute so that it provides an
important test to this approach.  The investigation is also extended to the
$\phi$ photoproduction, and the initial results show that the non-diffractive
behavior of the $\phi$ productions in the large $t$ region
can be described by the s- and u- channel
 contributions with significantly smaller coupling constants, which is 
consistent with the findings in the similar studies in the QHD framework.
The numerical investigation has also
shown that polarization observables are essential for  identifying
so-called ``missing resonances". 
\end{abstract}

\newpage
\section*{\bf 1. Introduction}
The newly established electron and photon facilities have made 
it possible to investigate the mechanism of vector meson 
photoproductions on nucleons with much improved experimental accuracy. 
This has been motivated in part by 
the puzzle that the NRQM ~\cite{NRQM,capstick} predicts 
a much richer resonance spectrum than has been observed 
in $\pi N\to \pi N$ scattering experiments. 
Quark model studies have suggested that those resonances 
missing in the $\pi N$ channel may couple strongly to, 
for example, the $\omega N$ and $\rho N$ channels. 
Experiments have been performed at ELSA~\cite{saphir} and will be done 
at TJNAF in the near future~\cite{cebaf}. 
Therefore, a theory on the reaction mechanism 
that highlights the dynamical role of 
s-channel resonances is crucial in order to establish the ``missing 
resonances" in the forthcoming experimental data for the vector meson, in
particular $\omega$ and $\rho$, photoproductions.

The experimental and theoretical studies on vector meson 
photoproductions have been a long history since the first
experiment carried out on the Cambridge Electron Accelerator
in 1964. There have been some experimental data of the vector meson 
photoproductions in their threshold regions where the s-channel
resonances are expected to play an important role, for instance,
the data from Ref.\cite{saphir,ABHMC,benz,hilpert} for the $\omega$ and 
$\rho$ photoproduction ($E^{thres}_\gamma\simeq$ 1.11GeV) , 
and the data from Ref.\cite{phidata} for the $\phi$ 
photoproduction ($E^{thres}_\gamma\simeq$ 1.57GeV). 
Historically, these studies have concentrated on the diffactive
behavior in the small $t$ region for the neutral meson ($\omega$, 
$\rho^0$ and $\phi$) productions, in which it has been shown that the 
Vector Meson Dominance Model (VMD)~\cite{Bauer} gives a good description 
in the low energy region, while the Pomeron exchange becomes more 
important as the energy increases\cite{pomeron}.  The focus of this 
paper is not on the large diffractive behavior in the small $t$ region, 
 but rather on the non-diffractive s- and u- channel contributions in 
the large $t$ region in which the t-channel Pomeron
exchange becomes less significant.  Furthermore, we will attempt to
provide a unified framework for both neutral and charged vector meson
productions. Since the t-channel exchanges responsible for the large
diffractive behavior in the small $t$ region, such as the Pomeron exchange,
does not contribute to the charged meson productions,  the non-diffractive 
s- and u-channel resonances become more dominant, thus they provide a 
crucial test to any model that concentrates on the role of the s- and 
u-channel resonances in the vector meson photoproductions. The quark model 
approach provides an ideal framework to investigate the dynamical role of 
the s- and u-channel resonances in the vector meson photoproductions. The
 studies in the pseudoscalar photoproductions have shown\cite{pseudoscalar} 
that every s- and u-channel resonance,  particularly those high partial 
wave resonances such as $F_{15}$ and $F_{37}$, can be taken into account, 
and this has been proven to be very difficult for the traditional approach 
at the hadronic level. Moreover, it introduces the quark degrees of freedom
 directly into the reaction mechanism, thus gives a very good description 
to the pseudoscalar meson productions with much less parameters than the
models at the hadronic level. It is therefore natural to extend this 
approach to the vector meson photoproductions in the resonance region.

A major difference for the vector meson production from the 
pseudoscalar meson case in the quark model is that the 
interaction between the vector mesons and the quarks inside 
the baryon is largely unknown.  Although phenomenological models 
have been developed to evaluate baryon resonance 
decaying into a nucleon and a vector meson,  
 such as the quark pair creaction model~\cite{yaouanc} 
or the $^3P_0$ model,
these approaches are unsuitable for the description of 
vector meson photoproductions. This is due to the fact that they
 only yield transition amplitudes for s-channel resonances, but
contain no information on how to derive the non-resonant terms 
in the u- and t-channels.  Therefore,  we choose an effective 
Lagrangian here as our starting point that satisfies the fundamental 
symmetries and determines the transition amplitudes not only 
in the s-channel but also in the u- and t- channels.

Even though the effective Lagrangians are different from each other for 
pseudoscalar and vector meson photoproductions, 
the implementation follows the same guidelines.
The transition amplitudes for each resonance in the s-channel
below 2 GeV will be included explicitly, 
while the resonances above 2 GeV for a given quantum number $n$ 
in the harmonic oscillator basis of the quark model
are treated as degenerate, so that their transition amplitudes can be written 
in a compact form. Similarly, the excited resonances in the u-channel are 
assumed to be degenerate as well. Only the mass splitting between the 
spin 1/2 and spin 3/2 resonances with $n=0$ in the harmonic oscillator basis,
such as the splitting between nucleon and $\Delta$ resonance, is
found to be significant, thus, 
the transition amplitudes for the spin 3/2 resonance with $n=0$ in 
the u-channel will be included separately.

The effective Lagrangian employed here generates not only the s- and u-channel
exchanges but also a t-channel term containing the vector meson exchange.
For charged vector mesons gauge invariance also mandates a Seagull term.
Although in principle, all the contributions from resonances have 
been included in the effective Lagrangian, we don't expect that 
such an approach can give an agreement with the data in the small 
$t$ region in the neutral vector meson photoproductions 
since additional t-channel contribution, such as the Pomeron exchange, 
will responsible for the strong diffractive behavior 
in the small $t$ region. As it has been shown
in Ref.\cite{collins} and also in Ref.\cite{Freund} and Ref.\cite{Harari}
about the diffraction duality, there can be a non-resonant 
imaginary background amplitude in neutral vector meson,
such as $\omega$ and $\rho^0$ photoproductions,
but not in the charged vector meson $\rho^\pm$ photoproductions.
Therefore, the large difference of the cross section 
between the neutral and charged $\rho$ meson photoproduction, 
observed  in the direct channel resonance region, 
is due to such a background amplitude, and it should be 
from the large contribution of the Pomeron singularity in the 
neutral photoproduction from high energies down to the threshold.
This has been one of our concerns in the numerical 
investigations.  Therefore, we add a t-channel $\pi^0$ exchange to the
amplitude  for $\omega$ photoproduction and a $\sigma$ exchange term to 
the amplitude for $\rho^0$ photoproduction
suggested by Friman and Soyeur~\cite{FrimanSoyeur} 
who showed  these two terms play  dominant roles in $\omega$ and $\rho^0$
productions respectively, over other meson exchange processes near the
threshold. 

With the above considerations, we apply our model to the five 
isospin channels, $\gamma p\to \omega p$, $\gamma p\to \rho^0 p$, 
$\gamma n\to \rho^- p$, $\gamma p\to \rho^+ n$ and
$\gamma p\to \phi p$.
With the same set of parameters introduced in our model, we 
obtained an overall agreement with the differential cross sections
in the large $t$ region for the first four channels,
while
with relatively smaller parameters in the $\phi$ photoproduction, 
we predict the behavior of the differential cross section in the 
large $t$ region.
With the additional t-channel $\pi^0$ and $\sigma$ exchanges
included in the $\omega$ and $\rho^0$ photoproduction respectively,
we obtain an overall agreement with the available data from 
small $t$ to large $t$ region.  
The overall agreement between the theoretical predictions and the data
available not only for the neutral meson $\omega$ and $\rho^0$ but also
for the charged meson $\rho^\pm$ productions in which the s- and u-channel
contributions become more dominant is remarkable. It is even more
remarkable that both $\omega$ and $\rho$ productions can be described by
the same set of parameters, which is by no means trivial. 
It suggests that
quark model approach provides a very good framework to investigate the
resonance structure in the vector meson photoproductions. Our results also
show that polarization observables is crucial in determining the resonance
structure, which has been shown to be the case in the pseudoscalar meson
photoproductions.

In the reaction $\gamma p\to \phi p$.  
Since the threshold energy of the $\phi$ production is above the resonance 
region, the primary focus here is the $\phi NN$ coupling constant. Because
 the production of $s\overline{s}$ from the nucleons should be suppressed
under the Okubo-Zweig-Iizuka (OZI) rule, the $\phi NN$ coupling constant
is expected to be smaller than the $\omega NN$ or $\rho NN$ couplings.
 The recent experiment on the 
$p\overline{p} \to \phi X$ ($X=\pi, \eta, \omega, \rho, \pi\pi, \gamma$)
\cite{OZI} has shown a significant violation of the OZI rule, and it
can not be explained by the diffractive process, such as Pomeron 
exchange. Thus, large contributions from the non-diffractive processes 
are expected to contribute to the $\phi$ photoproduction near the threshold. 
This has been the subject of many studies, such as the recently developed
 quantum hadrodynamical(QHD) model approach\cite{williams}.
In our framework, it could be
achieved by fitting the s- and u- channel contributions 
to the differential cross sections of $\phi$ productions in the large $t$ 
region\cite{phidata}, where the contributions from the Pomeron exchange 
become less significant.  The initial results show that the $\phi NN$ 
couplings in the quark model are small but significant which is consistent 
with those in the QHD approach. 

In Section 2, we briefly discuss some of the observables
used in our approach, which have been developed extensively in Ref. 
\cite{tabakin}.  The framework for the vector meson
photorpductions with effective Lagrangian for the quark-meson 
interaction is presented in Section 3.  In Section 4, we show our
numerical studies of the $\omega$, $\rho$ and $\phi$ 
photoproductions in the five isospin channels. 
Finally, conclusions will be presented in Section 5.
  
\section*{\bf 2. Observables and Helicity Amplitudes }  
  
Before presenting our quark model approach we 
introduce some general features
of vector meson photoproduction on the nucleon.
The basic amplitude $\cal F$ for $\gamma + N \to V+N^\prime$ is defined as
\begin{equation}  
{\cal F}=\langle {\bf q}\lambda_V\lambda_2|T|{\bf k}
 \lambda\lambda_1\rangle,  
\label{1}  
\end{equation}  
where ${\bf k}$ and ${\bf q}$ are the momenta of the incoming 
photon and outgoing vector meson.
The helicity states are denoted by $\lambda=\pm 1$ 
for the incident photon,   
$\lambda_V=0,\pm 1$ for the outgoing  vector meson, 
and $\lambda_1=\pm 1/2$,$\lambda_2=\pm 1/2$  
for the initial and final state nucleons, respectively. Following 
Ref.~\cite{tabakin}, the amplitude $\cal F$
can be expressed as a $6\times 4$ matrix in the helicity space,
\begin{equation}  
{\cal F}=\left( \begin{array}{cccc}  
H_{21} & H_{11} & H_{3-1} & -H_{4-1}\\  
H_{41} & H_{31} & -H_{1-1} & H_{2-1}\\
H_{20} & H_{10} & -H_{30} & H_{40}\\
H_{40} & H_{30} & H_{10} & -H_{20}\\
H_{2-1} & H_{1-1} & H_{31} & -H_{41}\\
H_{4-1} & H_{3-1} & -H_{11} & H_{21}\\ 
\end{array} \right ).
\label{2}
\end{equation}  
Because of the parity conservation,  
\begin{equation}  
\langle {\bf q}\lambda_V\lambda_2|T|{\bf k} \lambda\lambda_1\rangle=  
(-1)^{\Lambda_f-\Lambda_i}\langle {\bf q} -\lambda_V -\lambda_2|T|{\bf k}-  
\lambda -\lambda_1\rangle,  
\end{equation}
where $\Lambda_f=\lambda -\lambda_1$ and $\Lambda_i=\lambda_V -\lambda_2$
 in the Jacob-Wick(JW) convention,  
the $ H_{a\lambda_V}(\theta)$ in Eq.(\ref{2}) reduces to 12 independent  
 complex helicity amplitudes:

\begin{eqnarray} \label{helicity}
H_{1\lambda_V}&=  &\langle \lambda_V, \lambda_2=+1/2|T|\lambda=1,   
\lambda_1=-1/2\rangle\nonumber\\  
H_{2\lambda_V}&=  &\langle \lambda_V, \lambda_2=+1/2|T|\lambda=1,   
\lambda_1=+1/2\rangle\nonumber\\  
H_{3\lambda_V}&=  &\langle \lambda_V, \lambda_2=-1/2|T|\lambda=1,
\lambda_1=-1/2\rangle\nonumber\\  
H_{4\lambda_V}&=  &\langle \lambda_V, \lambda_2=-1/2|T|\lambda=1,  
\lambda_1=+1/2\rangle.  
\end{eqnarray}  

Each experimental observable $\Omega$ can be written in the general 
{\sl bilinear helicity product}(BHP) form,
\begin{eqnarray}
\check{\Omega}^{\alpha\beta}&=&\Omega^{\alpha\beta}{\cal T}(\theta)\nonumber\\
&=&\pm\frac 12\langle H|\Gamma^\alpha\omega^\beta|H \rangle\nonumber\\
&=&\pm\frac 12\sum_{a,b,\lambda_V,\lambda^\prime_V} 
H^*_{a\lambda_V}\Gamma^\alpha_{ab}\omega^\beta_{\lambda_V\lambda^\prime_V}
H_{b\lambda^\prime_V}.
\end{eqnarray}
For example, the differential cross 
section operator is given by:
\begin{eqnarray}
\check{\Omega}^{\alpha=1,\beta=1}&=&{\cal T}(\theta)\nonumber\\
&=&\frac 12\langle H|\fbox{$\Gamma^1$}\fbox{$\omega^1$}|H\rangle\nonumber\\
&=&\frac 12\sum^4_{a=1}\sum_{\lambda_V=0,\pm 1} |H_{a\lambda_V}|^2,
\end{eqnarray}
where the box
 frames denote the diagonal structure of the matrices. The
$\Gamma$ and $\omega$ matrices labeled by different $\alpha$ and $\beta$ 
correspond to different spin observables. With the phase space factor, the 
differential cross section has the expression,
\begin{eqnarray}
\frac{d\sigma}{d\Omega_{c.m.}}&=&(P.S. factor){\cal T}(\theta)\nonumber\\
&=&\frac{\alpha_e \omega_m(E_f+M_N)(E_i+M_N)}{8\pi s}{|{\bf q}|}\frac 12
\sum^4_{a=1}\sum_{\lambda_V=0,\pm 1} |H_{a\lambda_V}|^2
\end{eqnarray}
in the center of mass frame, where $\sqrt{s}$ is the total energy of the
 system, $E_i$ and $E_f$ are the energies of the nucleons in the 
initial and final states, respectively. 
$M_N$ represents the masses of the nucleon,  
and $\omega_m$ denotes the energy of the outgoing meson.

These helicity amplitudes are usually related to the density matrix 
elements $\rho_{ik}$~\cite{schilling}, which are measured 
by the experiments~\cite{Ballam}. 
They are defined as:  
\begin{eqnarray}  
\rho^0_{ik}&=  &\frac{1}{A}\sum_{\lambda\lambda_2\lambda_1}H_{\lambda_{V_i}  
\lambda_2,  
\lambda\lambda_1} H^*_{\lambda_{V_k}\lambda_2, \lambda\lambda_1},  
\nonumber\\
\rho^1_{ik}&=  &\frac{1}{A}\sum_{\lambda\lambda_2\lambda_1}  
H_{\lambda_{V_i}\lambda_2,  
-\lambda\lambda_1} H^*_{\lambda_{V_k}\lambda_2, \lambda\lambda_1},  
\nonumber\\  
\rho^2_{ik}&=  &\frac{i}{A}\sum_{\lambda\lambda_2\lambda_1}\lambda   
H_{\lambda_{V_i}\lambda_2, -\lambda\lambda_1} H^*_{\lambda_{V_k}\lambda_2,   
\lambda\lambda_1},\nonumber\\  
\rho^3_{ik}&=  &\frac{i}{A}\sum_{\lambda\lambda_2\lambda_1}\lambda   
H_{\lambda_{V_i}\lambda_2, \lambda\lambda_1} H^*_{\lambda_{V_k}\lambda_2,   
\lambda\lambda_1},   
\end{eqnarray}  
where   
\begin{equation}  
A=\sum_{\lambda_{V_i}\lambda\lambda_2\lambda_1} H_{\lambda_{V_i}\lambda_2,  
\lambda\lambda_1} H^*_{\lambda_{V_k}\lambda_2, \lambda\lambda_1},  
\end{equation}  
where $\rho_{ik}$ stands for $\rho_{\lambda_{V_i}\lambda_{V_k}}$,
 and $\lambda_{V_i}$, $\lambda_{V_k}$ denote the helicity of the 
produced vector mesons.

For example, the angular distribution  
 for $\rho^0$ decaying into $\pi^+\pi^-$ 
produced by linearly polarized photons can be expressed  
 in terms of nine independent measurable spin-density matrix elements
\begin{eqnarray}  
W(cos\theta, \phi,\Phi)&=  &\frac 3{4\pi}[\frac 12(1-\rho^0_{00})  
+\frac 12(3\rho^0_{00}-1)cos^2\theta-\sqrt2Re\rho^0_{10}sin2  
\theta cos\phi\nonumber\\
&&-\rho^0_{1-1}sin^2\theta cos2\phi  
-P_\gamma cos2\Phi(\rho^1_{11}sin^2\theta+\rho^1_{00}cos^2  
\theta\nonumber\\
&&-\sqrt2 Re\rho^1_{10}sin2\theta cos\phi-\rho^1_{1-1}sin^2  
\theta cos2\phi)\nonumber\\
&&-P_\gamma sin2\Phi(\sqrt2 Im\rho^2_{10}sin2\theta sin\phi
+Im\rho^2_{1-1}sin^2\theta sin2\phi)],  
\end{eqnarray}  
where $P_\gamma$ is the degree of the linear polarization of the  
 photon, $\Phi$ is the angle of the photon electric polarization  
 vector with respect to the production plane measured in the c.m.  
 system, and $\theta$ and $\phi$ are the polar and azimuthal angles   
of the $\pi^+$ which is produced by the $\rho^0$ decay in the   
$\rho^0$ rest frame.

\section*{\bf 3. Quark Model Approach for Vector Meson Photoproduction}  
  
The starting point of the quark model approach is the effective Lagrangian,
\begin{equation} \label{3.0}  
L_{eff}=-\overline{\psi}\gamma_\mu p^\mu\psi+\overline{\psi}\gamma_
\mu e_qA^\mu\psi +\overline{\psi}(a\gamma_\mu +  
\frac{ib\sigma_{\mu\nu}q^\nu}{2m_q}) \phi^\mu_m \psi,
\end{equation}  
where the quark field $\psi$ is expressed as  
\begin{equation}  
\psi =\left( \begin{array}{c}  
\psi (u)\\ \psi (d) \\ \psi (s) \end{array} \right ),   
\end{equation}  
and the meson field $\phi^\mu_m $ is a 3$\otimes$3 matrix,  
\begin{equation}  
\phi_m =\left( \begin{array}{ccc}  
\frac{1}{\sqrt{2}}\rho^{0}+\frac{1}{\sqrt{2}}\omega & \rho^{+} & K^{*+}\\  
\rho^{-} & -\frac{1}{\sqrt{2}}\rho^{0}+\frac{1}{\sqrt{2}}\omega & K^{*0}\\  
K^{*-} & \overline{K}^{*0}  &\phi  
\end{array} \right )   
\end{equation}  
in which the vector mesons are treated as point-like particles. 
At tree level, the transition matrix element based on the effective   
Lagrangian in Eq.(\ref{3.0}) can be written as the sum of contributions  
 from the s-, u- and t- channels,
\begin{equation}  
M_{fi}=M^s_{fi}+M^u_{fi}+M^t_{fi},  
\label{3.1}  
\end{equation}  
where the s- and u-channel contributions in Eq.(\ref{3.1}) 
 have the following form,  
\begin{eqnarray}  
M^s_{fi}+M^u_{fi}&=  &\sum_{j} \langle N_f|H_m|N_j\rangle\langle   
N_j|\frac{1}{E_i+\omega-E_j}H_e|N_i\rangle\nonumber\\  
&&+\sum_{j} \langle N_f|H_e\frac{1}{E_i-\omega_m-E_j}
|N_j\rangle\langle N_j|H_m|N_i\rangle,  
\end{eqnarray}  
where the electromagnetic coupling vertex is  
\begin{equation}  
H_e=-\overline{\psi}\gamma_\mu e_qA^\mu\psi,  
\end{equation}  
and the quark-meson coupling vertex is  
\begin{equation} \label{Hm} 
H_m=-\overline{\psi}(a\gamma_\mu +\frac{ib\sigma_{\mu\nu}  
q^\nu}{2m_q}) \phi^\mu_m \psi,  
\end{equation}  
where $m_q$ in is the quark mass and the constants $a$ 
and $b$ in Eq.(\ref{3.0})
and (\ref{Hm}) are the vector and tensor   
coupling constants, which will be treated as free 
parameters in our approach. The initial and final
states of the nucleon are denoted by $|N_i\rangle$ and $|N_f\rangle$, 
respectively, and $|N_j\rangle$
 is the intermediate resonance state while $E_i$ and $E_j$ 
are the energies of the inital nucleon and the intermediate resonance. 

An important test of the transition matrix elements 
\begin{equation}  
M_{fi}=\langle \lambda_2|J_{\mu\nu}\epsilon^\mu
\epsilon^\nu_m|\lambda_1\rangle,  
\label{2.1}  
\end{equation}  
would be gauge invariance:  
\begin{equation}  
\langle \lambda_2|J_{\mu\nu}k^\mu|\lambda_1\rangle=\langle  
 \lambda_2|J_{\mu\nu} q^\nu|\lambda_1\rangle=0,  
\label{2.2}  
\end{equation}  
where $\epsilon^\mu_m$, $\epsilon^\nu$ are the polarization  
 vectors of the vector mesons and photons.
However, we find that the condition $\langle  
 \lambda_2|J_{\mu\nu} q^\nu|\lambda_1\rangle=0$ is not satisfied for
the t-channel vector meson exchange term,
\begin{equation}  
M^t_{fi}=-a\langle N_f|\sum_{l}\frac{e_m}{2q\cdot k}  
\{2q\cdot\epsilon\gamma\cdot\epsilon_m-\gamma\cdot   
q\epsilon\cdot\epsilon_m+k\cdot\epsilon_m\gamma\cdot\epsilon\}  
e^{i({\bf k}-{\bf q})\cdot{\bf r}_l}| N_i\rangle,  
\label{3.3}  
\end{equation} 
based on the Feynman rules for the photon-vector meson coupling, and
the relation
\begin{equation}  
\langle N_f|\gamma\cdot(q-k) e^{i({\bf k}-{\bf q})  
\cdot{\bf r}_l}| N_i\rangle=0.  
\end{equation}  
To remedy this problem, we add a gauge fixing term, so that 
\begin{equation}  
M^t_{fi}=-a\langle N_f|\sum_{l}\frac{e_m}{2q\cdot k}  
2\{q\cdot\epsilon\gamma\cdot\epsilon_m-\gamma\cdot q\epsilon\cdot\epsilon_m+k\cdot\epsilon_m\gamma\cdot\epsilon\}  
e^{i({\bf k}-{\bf q})\cdot{\bf r}_l}| N_i\rangle.  
\end{equation}

The techniques of deriving the transition amplitudes have been developed
for Compton scattering~\cite{Compton}. Follow the same procedure as given 
in Eq.(14) of Ref.~\cite{pseudoscalar}, we can divide the photon 
interaction into two parts and the contributions 
from the s- and u- channel can be rewritten as:
\begin{eqnarray}  
M^{s+u}_{fi}&=&i\langle N_f|[g_e,H_m]|N_i\rangle \nonumber\\  
&&+i\omega\sum_{j}\langle N_f|H_m|N_j\rangle\langle   
N_j|\frac{1}{E_i+\omega-E_j}h_e|N_i\rangle\nonumber\\  
&&+i\omega\sum_{j} \langle N_f|h_e\frac{1}{E_i-\omega_m-E_j}  
|N_j\rangle\langle N_j|H_m|N_i\rangle, 
\label{3.2}  
\end{eqnarray}  
where  
\begin{equation}  
g_e=\sum_{l}e_l{{\bf r}_l\cdot\veps}e^{i{\bf k\cdot r}_l},  
\end{equation}  
\begin{equation}  
h_e=\sum_{l}e_l{{\bf r}_l\cdot\veps}(1-\valf\cdot  
{\hat{\bf k}})e^{i{\bf k\cdot r}_l}  
\end{equation}  
\begin{equation}
{\bf{\hat k}}=\frac{\bf k}{\omega}.
\end{equation}
 
The first term in Eq.(\ref{3.2}) can be identified as a seagull term;
it is proportional to the charge of the outgoing vector meson.
The second and third term in Eq.(\ref{3.2}) represents the s- and u-channel
contributions. Adopting the same strategy as in the pseudoscalar case,
 we include a complete set of helicity amplitudes for each
of the s-channel resonances below 2GeV in the $SU(6)\otimes O(3)$
 symmetry limit. The resonances above 2GeV
are treated as degenerate in order to express each contribution from  
all resonances with quantum number $n$ in a compact form.
The contributions from the resonances with the largest spin for  
 a given quantum number $n$ were found to be  the most 
important as the energy   
increases~\cite{pseudoscalar}.   
This corresponds to spin $J=n+1/2$ with $I=1/2$ for the reactions  
 $\gamma N\to K^*\Lambda$ and $\gamma N\to \omega N$,  
 and $J=n+3/2$ with $I=3/2$ for the reactions $\gamma N\to K^*\Sigma$ and  
 $\gamma N\to \rho N$.

Similar to the pseudoscalar case, the contributions from   
the u-channel resonances are divided into two parts as well. The
first part contains the resonances with   
the quantum number $n=0$, which includes the spin 1/2 states,  
 such as the $\Lambda$, $\Sigma$ and the nucleons, 
and the spin 3/2 resonances, such as the $\Sigma^*$ in $K^*$ photoproduction
and $\Delta(1232)$ resonance in $\rho$ photoproduction. Because   
the mass splitting between the spin 1/2 and spin 3/2 resonances   
for $n=0$ is significant, they have to be treated separately.   
The transition amplitudes for these u-channel resonances will  
also be written in terms of the helicity amplitudes.
 The second part comes from the excited   
resonances with quantum number $n\ge 1$. As the contributions  
 from the u-channel resonances are not sensitive to the precise 
mass positions, they can be treated as degenerate as well,
so that the contributions from these resonances can again
be written in a compact form.  

\subsection*{\bf 3.1. The Seagull term }  

The transition amplitude is divided into the transverse
and longitudinal amplitudes according to the polarization of the outgoing 
vector mesons. The longitudinal polarization vector for a vector meson with
mass $\mu$ and momentum ${\bf q}$ is,
\begin{equation}    
\epsilon^\mu_L =\frac{1}{\mu} \left( \begin{array}{c}    
|{\bf q}|\\ \omega_m \frac{\bf q}{|{\bf q}|} \end{array} \right )     
\end{equation}    
where $\omega_m=\sqrt{{\bf q}^2+\mu^2}$ is the energy of the outgoing 
vector mesons.
Thus, the longitudinal interaction at the quark-meson vertex 
can be written as    
\begin{equation}    
H^L_m=\epsilon^\mu_L J_\mu=\epsilon_0J_0-\epsilon_3J_3    
\end{equation}    
where $\epsilon_3$ corresponds to the direction of the momentum ${\bf q}$.
The transition amplitudes of the s- and u-channel for the 
longitudinal quark-meson coupling become,    
\begin{eqnarray}    
M^{s+u}_{fi}(L)&=& i\langle N_f|[g_e,H^L_m]|N_i\rangle \nonumber \\    
&&- i\omega \langle N_f|[h_e,\frac{\epsilon_3}{q_3}J_0]|N_i\rangle  
\nonumber \\    
&&+i\omega\sum_{j}\langle N_f|(\epsilon_0
-\frac{\omega_m}{q_3}\epsilon_3)J_0|N_j\rangle
\langle N_j|\frac{1}{E_i+\omega-E_j}h_e|N_i\rangle\nonumber \\    
&&+i\omega\sum_{j} \langle N_f|h_e\frac{1}{E_i-\omega_m-E_j}|N_j  
\rangle\langle N_j|(\epsilon_0-\frac{\omega_m}{q_3}\epsilon_3)J_0|N_i
\rangle,    
\label{4.2})   
\end{eqnarray}    
where the first two terms are seagull terms which can be rewritten
as,
\begin{equation}
\label{seagull}
M^{Seagull}_{fi}(L)=-\frac{a\omega_m e_m}{\mu |{\bf q}|}\langle N_f|\valf\cdot\veps e^{i({\bf k}-{\bf q})\cdot{\bf r}_l}|N_i\rangle -\frac{ia\mu e_m}{|{\bf q}|}\langle N_f|\sum_l {\bf r}_l\cdot\veps e^{i({\bf k}-{\bf q})\cdot{\bf r}_l}|N_i\rangle .
\end{equation}
The first term will be cancelled by a corresponding term from 
the t-channel to keep the gauge invariance while the second term 
has more explicit expression as,
\begin{equation}
\label{seagull-longi}
M^{Seagull}_{fi}(L)=-\frac{a\mu e_m}{|{\bf q}|\alpha^2}g^t_v {\bf q}
\cdot\veps e^{i({\bf k}-{\bf q})\cdot{\bf r}_l},
\end{equation}
where the $g$- factor $g^t_v$ is defined as the following,
\begin{equation}
\label{gtv}
g^t_v=\langle N_f|\sum_j{\hat I}_j|N_i\rangle.
\end{equation}
The values of $g^t_v$ for every channel is presented in Table 1.
The corresponding expressions for the t-channel amplitudes
are given in the Appendix. The last two terms in (\ref{4.2}) 
will be discussed in the following sections.

The nonrelativistic expansion of the transverse meson quark
interaction vertex gives,    
\begin{equation}    
H^T_m=\sum_{l} \{i\frac{b^\prime}{2m_q}\vsig_l\cdot({\bf q}
\times\veps_v)  
+a{\bf A}\cdot\veps_v+\frac{a}{2\mu_q}{\bf p}^\prime_l  
\cdot\veps_v\}{\hat I}_le^{-i{\bf q\cdot r}_l}     
\end{equation}    
where $b^\prime\equiv b-a$, ${\bf p}^\prime_l$ is the internal 
motion of the $l$th quark in the c.m. system, and 
\begin{equation}  
{\hat I}_l =\left\{ \begin{array}{ccc}
a^\dagger _l(s)a_l(u) & \qquad\mbox{for}\qquad & K^{*+}\\ 
a^\dagger_l(s)a_l(d) & \qquad\mbox{for}\qquad & K^{*0}\\
a^\dagger_l(d)a_l(u) & \qquad\mbox{for}\qquad & \rho^+\\
-\frac{1}{\sqrt 2}(a^\dagger _l(u)a_l(u) 
-a^\dagger _l(d)a_l(d)) & \qquad\mbox{for}\qquad & \rho^0\\
1 &\qquad\mbox{for}\qquad & \omega (\phi )
\end{array} \right.   
\end{equation}  
The vector ${\bf A}$ has the general form, 
\begin{equation}    
{\bf A}=\frac{{\bf P}_f}{E_f+M_f}+\frac{{\bf P}_i}{E_i+M_i},    
\end{equation}    
which comes from the center-mass motion of the quark system. In   
the s- and u-channel, ${\bf A}$ has the following expression for  
 different channels,    
\begin{eqnarray}    
s-channel:&&{\bf A}=-\frac{{\bf q}}{E_f+M_f},\\    
u-channel:&&{\bf A}=-(\frac{1}{E_f+M_f}
+\frac{1}{E_i+M_i}){\bf k}-\frac{1}  
{E_f+M_f}{\bf q}.      
\end{eqnarray}    
    
The transverse transition amplitude for the s- and u-channel is,    
\begin{eqnarray}    
M^{s+u}_{fi}(T)&=&i\langle N_f|[g_e,H^T_m]|N_i\rangle \nonumber\\    
&&+i\omega\sum_{j}\langle N_f|H^T_m|N_j\rangle\langle N_j|
\frac{1}{E_i+\omega-E_j}h_e|N_i\rangle\nonumber\\
&&+i\omega\sum_{j} \langle N_f|h_e\frac{1}{E_i-\omega_m-E_j}|N_j  
\rangle\langle N_j|H^T_m|N_i\rangle    
\label{4.3}    
\end{eqnarray}    
The nonrelativistic expansion of the first term gives,
\begin{eqnarray}
M^{Seagull}_{fi}(T)&=&-i\langle N_f|[g_e, H^T_m]|N_i\rangle\nonumber\\
&=&-iae_m g^t_v\langle N_f|\{ \frac{{\bf P}\cdot\veps_v}{E+M}, 
{\bf R}\cdot\veps \}e^{-\frac{({\bf k-q})^2}{6\alpha^2}} |N_i\rangle\nonumber\\
&&+ae_m g_A\langle N_f| [\frac{\vsig\cdot({\bf P}\times\veps_v)}{E+M},
 {\bf R}\cdot\veps] e^{-\frac{({\bf k-q})^2}{6\alpha^2}}|N_i\rangle,
\end{eqnarray}
where $\{A,B\}=AB+BA$ is the anti-commutation operator. 
${\bf P}$ and ${\bf R}$ 
are the momentum and coordinate of the center mass 
motion of the three quark system. 

The seagull terms in the  
 transitions are  proportional to the charge of the 
outgoing mesons and, therefore,  
 vanish in the neutral vector meson, $\omega$, 
$\rho^0$ and $\phi$, photoproductions. 

\subsection*{\bf 3.2. U-channel transition amplitudes}    

The last term in Eq.(\ref{4.2}) is the longitudinal transition amplitude  
 in the u-channel. We find
\begin{eqnarray}    
M^u_{fi}(L)&=&i\omega\sum_{j} 
\langle N_f|h_e\frac{1}{E_i-\omega_m-E_j}|N_j\rangle\langle N_j  
|-\frac{\mu}{|{\bf q}|}J_0|N_i\rangle\nonumber\\    
&=& (M^u_3+M^u_2)e^{-\frac{{\bf q}^2+{\bf k}^2}{6\alpha^2}}    
\end{eqnarray}    
in the harmonic oscillator basis, where    
\begin{eqnarray}    
M^u_3&=&g^u_3 \frac{a\mu}{|{\bf q}|}
\{\frac{i}{2m_q}\vsig\cdot(\veps\times{\bf k})F^0  
(\frac{{\bf k}\cdot {\bf q}}{3\alpha^2},P_f\cdot k)\nonumber\\    
&&-g_v\frac{\omega}{3\alpha^2}{\bf q}\cdot\veps F^1  
(\frac{{\bf k}\cdot {\bf q}}{3\alpha^2},P_f\cdot k)\},    
\label{5.2}    
\end{eqnarray}    
which corresponds to incoming photons and outgoing vector mesons 
being absorbed and emitted by the same quark, and    
\begin{eqnarray}    
M^u_2&=&g^u_2 \frac{a\mu}{|{\bf q}|}
\{g^\prime_a\frac{i}{2m_q}\vsig\cdot(\veps\times{\bf k})  
F^0(-\frac{{\bf k}\cdot {\bf q}}{6\alpha^2},P_f\cdot k)\nonumber\\    
&&+g^\prime_v\frac{\omega}{6\alpha^2}{\bf q}\cdot\veps   
F^1(-\frac{{\bf k}\cdot {\bf q}}{6\alpha^2},P_f\cdot k)\}    
\label{5.3}
\end{eqnarray}    
in which the incoming photons and outgoing vector mesons are absorbed and
emitted by different quarks. $P_f$ in Eq.(\ref{5.2}) and Eq.(\ref{5.3}) 
denotes the four momentum of the final state nucleon.
 The function $F$ in Eq.(\ref{5.2}) and Eq.(\ref{5.3}) is defined as,    
\begin{equation}    
F^l(x,y)=\sum_{n\ge l}\frac{M_n}{(n-l)!(y+n\delta M^2)}x^{n-l},    
\end{equation}    
where $n\delta M^2=(M^2_n-M^2_f)/2$ represents 
the average mass difference between the ground state and excited states 
with the total excitation quantum number $n$ in the harmonic 
oscillator basis. The parameter $\alpha^2$ in the above equation is commonly 
used in the quark model and is related to the harmonic oscillator strength.   
    
Similarly, the transverse transition in the u-channel is given by,
\begin{eqnarray}    
M^u_{fi}(T)&=&i\omega\sum_{j} \langle N_f|
h_e\frac{1}{E_i-\omega_m-E_j}|N_j\rangle\langle N_j|H^T_m|N_i\rangle\nonumber\\    
&=& (M^u_3+M^u_2)e^{-\frac{{\bf q}^2+{\bf k}^2}{6\alpha^2}}    
\end{eqnarray}    
where     
\begin{eqnarray}    
M^u_3/g^u_3&=&\frac{b^\prime}{4m^2_q}\{g_v(\veps\times
{\bf k})\cdot({\bf   
q}\times\veps_v) +i\vsig\cdot(\veps\times{\bf k})\times({\bf q}  
\times\veps_v) \}F^0(\frac{{\bf k}\cdot {\bf q}}{3\alpha^2},P_f\cdot k)  
\nonumber\\    
&&-\frac{ia}{2m_q}\vsig\cdot(\veps\times{\bf k}){\bf A}\cdot\veps_v   
F^0(\frac{{\bf k}\cdot {\bf q}}{3\alpha^2},P_f\cdot k)\nonumber\\    
&&+\{\frac{ia}{12m^2_q}\vsig\cdot(\veps\times{\bf k})\veps_v\cdot{\bf   
k}+\frac{ib^\prime\omega}{6m_q\alpha^2}\vsig\cdot({\bf q}\times\veps_v)  
\veps\cdot{\bf q}\nonumber\\    
&&+g_v\frac{a\omega}{3\alpha^2}\veps \cdot{\bf q}{\bf A}\cdot\veps_v-   
g_v\frac{a\omega}{6m_q}\veps \cdot\veps_v \} F^1(\frac{{\bf k\cdot   
q}}{3\alpha^2},P_f\cdot k)\nonumber\\    
&&-g_v\frac{a\omega}{18m_q\alpha^2}\veps_v\cdot{\bf k}\veps  
\cdot{\bf q} F^2(\frac{{\bf k}\cdot {\bf q}}{3\alpha^2},P_f\cdot k)    
\label{amp3}    
\end{eqnarray}    
and    
\begin{eqnarray}    
M^u_2/g^u_2&= &\frac{b^\prime}{4m^2_q}\{g^\prime_v(\veps\times{\bf k})  
\cdot({\bf  
 q}\times\veps_v)\nonumber\\
&& -ig^\prime_a\vsig\cdot(\veps\times{\bf k})
\times({\bf q}\times\veps_v) \}F^0(-\frac{{\bf k}\cdot {\bf q}}  
{6\alpha^2},P_f\cdot k)  
\nonumber\\    
&&-\frac{ia}{2m_q}\vsig\cdot(\veps\times{\bf k}){\bf A}\cdot\veps_v   
F^0(-\frac{{\bf k}\cdot {\bf q}}{6\alpha^2},P_f\cdot k)\nonumber\\    
&&+\{-\frac{ia}{24m^2_q}\vsig\cdot(\veps\times{\bf k})\veps_v\cdot{\bf   
k}-\frac{ib^\prime\omega}{12m_q\alpha^2}\vsig\cdot({\bf q}\times\veps_v)  
\veps\cdot{\bf q}\nonumber\\    
&&-g^\prime_v\frac{a\omega}{6\alpha^2}\veps \cdot{\bf q}{\bf A}  
\cdot\veps_v- g^\prime_v\frac{a\omega}{12m_q}\veps \cdot\veps_v \}   
F^1(-\frac{{\bf k}\cdot {\bf q}}{6\alpha^2},P_f\cdot k)\nonumber\\    
&&-g^\prime_v\frac{a\omega}{72m_q\alpha^2}\veps_v\cdot{\bf k}  
\veps\cdot{\bf q} F^2(-\frac{{\bf k}\cdot {\bf q}}{6\alpha^2},P_f\cdot k)    
\label{amp2}    
\end{eqnarray}    
The $g$-factors in Eq.(\ref{5.2})-(\ref{amp2})
are defined as     
\begin{equation}     
\langle N_f|\sum_{j} {\hat I}_j \vsig_j|N_i\rangle=g_A \langle N_f|   
\vsig|N_i\rangle,    
\end{equation}      
\begin{equation} g^u_3=\langle N_f|\sum_{j} e_j {\hat
I}_j\sigma^z_j|N_i\rangle/g_A,    
\end{equation}    
\begin{equation} g^u_2=\langle N_f|\sum_{i\neq j} e_j {\hat   
I}_i\sigma^z_j|N_i\rangle/g_A,    
\end{equation}    
\begin{equation} g_v=\langle N_f|\sum_{j} e_j{\hat I}_j  
|N_i\rangle/g^u_3g_A,    
\end{equation}       
\begin{equation}    
g^\prime_v=\frac{1}{3g^u_2g_A}\langle N_f|\sum_{i\neq j}  
 {\hat I}_ie_j\vsig_i\cdot\vsig_j|N_i\rangle,    
\end{equation}    
\begin{equation}    
g^\prime_a=\frac{1}{2g^u_2g_A}\langle N_f|\sum_{i\neq j}   
{\hat I}_ie_j(\vsig_i\times\vsig_j)_z|N_i\rangle.    
\end{equation}    
The numerical values of these $g$-factors have been derived in 
Ref.~\cite{pseudoscalar} in the $SU(6)\otimes O(3)$ symmetry limit;
they are listed in Table 1 for completeness. 

The first terms of Eq.(\ref{amp3}) and Eq.(\ref{amp2}) correspond  
 to the correlation between the magnetic transition and the c.m.
motion of the meson transition operator and they contribute to the  
 leading Born term in the u-channel. The second terms are due to
 correlations between the internal and c.m. motion of the photon   
and meson transition operators, and they only contribute to the  
transitions between the 
 ground and $n\ge 1$ excited states in the harmonic oscillator  
 basis. The last terms in both equations represent the correlation
 of the internal motion between the photon and meson transition   
operators, which only contribute to transitions between the  
 ground and $n\ge 2$ excited states. 

As pointed out before, the mass splitting between 
the ground state spin 1/2 and spin 3/2  is significant, the transition
amplitudes for $\Delta$ resonance in $\rho$ production or $\Sigma^*$
resonance in $K^*$ production have to be computed separately.  
The transition amplitude with $n=0$ corresponding to the correlation
of magnitic transitions is,    
\begin{eqnarray}    
M^u(n=0)&=&-\frac{1}{2m_q}\frac{M_f e^{-\frac{{\bf q}^2+{\bf k}^2}
{6\alpha^2}}}{P_f\cdot k+\delta M^2/2}    
\frac{b^\prime}{2m_q}\{(g^u_3g_v+g^u_2g^\prime_v)({\bf k}\times\veps)  
\cdot({\bf q}\times\veps_v)\nonumber\\    
&&+i(g^u_3-g^u_2g^\prime_a)\vsig\cdot({\bf k}\times\veps)
\times({\bf q}\times\veps_v)\}.    
\end{eqnarray}    
The amplitude for spin 1/2 intermediate states in the total $n=0$  
 amplitudes is,    
\begin{eqnarray}    
&  &\langle N_f|h_e |N(J=1/2)\rangle\langle N(J=1/2)|H_m|N_i \rangle  
\nonumber\\    
&=  &\frac{\mu_N b^\prime}{2m_q}\frac{M_f
e^{-\frac{{\bf q}^2+{\bf k}^2}{6\alpha^2}}}  
{P_f\cdot k+\delta M^2/2}    
\{ ({\bf k}\times\veps)\cdot({\bf q}\times\veps_v)\nonumber\\    
&&+i\vsig\cdot({\bf k}\times\veps)\times({\bf q}\times\veps_v)\}    
\end{eqnarray}    
 where $\mu_N$ is the magnetic moment, which has the following   
values for different processes,    
\begin{equation}    
\mu_N =\left\{ \begin{array}{ccc}    
\mu_\Lambda+\frac{g_{K^*\Sigma N}}{g_{K^*\Lambda N}}\mu_{\Lambda\Sigma}   
  &\qquad\mbox{for}\qquad  & \gamma N  
\to K^*\Lambda\\    
\mu_{\Sigma^0}+\frac{g_{K^*\Lambda N}}{g_{K^*\Sigma N}}  
\mu_{\Lambda\Sigma}    &\qquad\mbox{for}\qquad &  
 \gamma N\to K^*\Sigma \\
\mu_{N_f}& \qquad\mbox{for}\qquad    &\gamma N  
\to \rho N_f     
\end{array} \right.    
\end{equation}    

Thus, we obtain the spin 3/2 resonance contribution to the transition   
amplitude by subtracting the spin 1/2 intermediate state contributions  
 from the total $n=0$ amplitudes as follows:
\begin{eqnarray}    
M^u&= &-\frac{b^\prime}{2m_q}\frac{M_fe^{-\frac{{\bf q}^2+{\bf k}^2}{6\alpha^2}}}  
{P_f\cdot k+\delta M^2/2}    
\{[(g^u_3g_v+g^u_2g^\prime_v)/2m_q-\mu_N]({\bf k}\times\veps)  
\cdot({\bf q}\times\veps_v)\nonumber\\    
&&+i[(g^u_3-g^u_2g^\prime_a)/2m_q-\mu_N]\vsig\cdot[({\bf   
k}\times\veps)\times({\bf q}\times\veps_v)]\}.
\end{eqnarray}    
Substituting the $g$-factor coefficients into the above equation 
gives the following general expression for spin 3/2 resonance with $n=0$,
\begin{eqnarray}    
M^u&= &-\frac{b^\prime}{2m_q}\frac{M_fg_se^{-\frac{{\bf q}^2
+{\bf k}^2}{6\alpha^2}}}  
{M_N(P_f\cdot k+\delta M^2/2)}
\{2\vsig\cdot({\bf q}\times\veps_v)\vsig\cdot({\bf k}\times\veps)\nonumber\\
&&-i\vsig\cdot[({\bf q}\times\veps_v)\times({\bf k}\times\veps)]\}    
\end{eqnarray}
where the value of $g_s$ is given in Table 1.

Note that the transition amplitudes here are generally written as 
operators that are similar to the CGLN amplitudes in pseudoscalar 
meson photoproduction. They have to be transformed into the helicity 
amplitudes defined in Eq.(\ref{helicity}). In Tables 2 and 3,
 we show the relations between the operators presented here and 
the helicity amplitudes; they
are generally related by the Wigner $d$-function. 
  
\subsection*{\bf 3.3 S-channel transition amplitudes }    
    
The third term in Eq.(\ref{4.2}) and second term in Eq.(\ref{4.3})  
 are the s-channel longitudinal and transverse transition amplitudes.   
 Following the derivation for Compton scattering  
 in Ref.~\cite{Compton}, we obtain the general transition amplitude for 
excited states in the s-channel,    
\begin{equation}  
H^J_{a\lambda_V}=\frac{2M_R}
{s-M_R(M_R-i\Gamma({\bf q}))}   
h^J_{a\lambda_V},    
\label{6.1}    
\end{equation}    
where $\sqrt{s}=E_i+\omega=E_f+\omega_m$ is the total energy   
of the system, and $H^J_{a\lambda_V}$ are the helicity amplitudes
defined previously. $\Gamma({\bf q})$ in Eq. \ref{6.1} denotes the  
 total width of the resonance,   which is
a function of the final state momentum ${\bf q}$.  For a resonance   
decaying into a two-body final state with relative angular momentum $l$,  
the decay width $\Gamma({\bf q})$ is  given by:
\begin{equation}\label{40}  
\Gamma({\bf q})= \Gamma_R \frac {\sqrt {s}}{M_R} \sum_{i} x_i   
\left (\frac {|{\bf q}_i|}{|{\bf q}^R_i|}\right )^{2l+1}   
\frac {D_l({\bf q}_i)}{D_l({\bf q}^R_i)},  
\end{equation}  
with   
\begin{equation}\label{41}  
|{\bf q}^R_i|=\sqrt{\frac   
{(M_R^2-M_N^2+M_i^2)^2}{4M_R^2}-M_i^2},  
\end{equation}  
and
\begin{equation}\label{42}  
|{\bf q}_i|=\sqrt{\frac   
{(s-M_N^2+M_i^2)^2}{4s}-M_i^2},  
\end{equation}  
where $x_i$ is the branching ratio of the resonance decaying into a   
meson with mass $M_i$ and a nucleon, and $\Gamma_R$ is the total   
decay width   
of the resonance with the mass $M_R$.  The   
function $D_l({\bf q})$ in Eq. \ref{40}, called fission barrier~\cite{bw}, 
is wavefunction dependent and has the following form
in the harmonic oscillator basis: 
\begin{equation}\label{43}  
D_l({\bf q})=exp\left (-\frac {{\bf q}^2}{3\alpha^2}\right ),  
\end{equation}  
which is independent of $l$. In principle, the branching ratio  
$x_i$ should also be evaluated in the quark model.    

For a given intermediate resonance state with spin $J$, the twelve   
independent helicity amplitudes $h^J_{a\lambda_V}$ in  Eq.(\ref{6.1})  
are a combination of the meson and photon helicity amplitudes  
 together with the Wigner-$d$ functions  
\begin{equation}
h^J_{a\lambda_V}=\sum_{\Lambda_f}d^J_{\Lambda_f,\Lambda_i}(\theta)  
A^V_{\Lambda_f}A^\gamma_{\Lambda_i},    
\label{59}
\end{equation}    
where $\Lambda_f=\lambda_V-\lambda_2$, $\Lambda_i=\lambda-\lambda_1$ 
and ${\bf k}\cdot{\bf q}=|{\bf k}||{\bf q}|cos(\theta)$.

The $A^\gamma_{1/2}$ and $A^\gamma_{3/2}$ in Eq.(\ref{59})  
represent the helicity amplitudes in the s-channel for the photon  
interactions; their explicit expressions have been given 
in Ref.~\cite{thesis}.

More explicitly, the 12 independent helicity amplitudes are related to
 the photon helicity amplitudes $A^\gamma_{\frac 12}$, $A^\gamma_{\frac 32}$ 
and vector meson helicity amplitudes $S^V_{\frac 12}$, 
$A^V_{\frac 12}$ and $A^V_{\frac 32}$ through the following relations

\begin{eqnarray}    
h^J_{11}&=&d^J_{\frac{1}{2},\frac{3}{2}}(\theta)A^V_{\frac{1}{2}}  
A^\gamma_{\frac{3}{2}},\nonumber\\    
h^J_{10}&=&d^J_{-\frac{1}{2},\frac{3}{2}}(\theta)S^V_{-\frac{1}{2}}
A^\gamma_{\frac{3}{2}},\nonumber\\    
h^J_{1-1}&=&d^J_{-\frac{3}{2},\frac{3}{2}}(\theta)A^V_{-\frac{3}{2}}    
A^\gamma_{\frac{3}{2}}
\end{eqnarray}    
for $a=1$, and $\lambda_V=1,0,-1$,
\begin{eqnarray}    
h^J_{21}&=&d^J_{\frac{1}{2},\frac{1}{2}}(\theta)A^V_{\frac{1}{2}}  
A^\gamma_{\frac{1}{2}},\nonumber\\    
h^J_{20}&=&d^J_{-\frac{1}{2},\frac{1}{2}}(\theta)S^V_{-\frac{1}{2}}    
A^\gamma_{\frac{1}{2}},\nonumber\\    
h^J_{2-1}&=&d^J_{-\frac{3}{2},\frac{1}{2}}(\theta)A^V_{-\frac{3}{2}}    
A^\gamma_{\frac{1}{2}}   
\end{eqnarray}    
for $a=2$, and $\lambda_V=1,0,-1$, 
\begin{eqnarray}    
h^J_{31}&=&d^J_{\frac{3}{2},\frac{3}{2}}(\theta)A^V_{\frac{3}{2}}    
A^\gamma_{\frac{3}{2}},\nonumber\\    
h^J_{30}&=&d^J_{\frac{1}{2},\frac{3}{2}}(\theta)S^V_{\frac{1}{2}}    
A^\gamma_{\frac{3}{2}},\nonumber\\    
h^J_{3-1}&=&d^J_{-\frac{1}{2},\frac{3}{2}}(\theta)A^V_{-\frac{1}{2}}    
A^\gamma_{\frac{3}{2}}  
\end{eqnarray}
for $a=3$, and $\lambda_V=1,0,-1$,
and 
\begin{eqnarray}    
h^J_{41}&=&d^J_{\frac{3}{2},\frac{1}{2}}(\theta)A^V_{\frac{3}{2}}   
A^\gamma_{\frac{1}{2}},\nonumber\\   
h^J_{40}&=&d^J_{\frac{1}{2},\frac{1}{2}}(\theta)S^V_{\frac{1}{2}}    
A^\gamma_{\frac{1}{2}},\nonumber\\    
h^J_{4-1}&=&d^J_{-\frac{1}{2},\frac{1}{2}}(\theta)A^V_{-\frac{1}{2}}    
A^\gamma_{\frac{1}{2}}   
\end{eqnarray}    
for $a=4$, and $\lambda_V=1,0,-1$.

The amplitudes with negative helicities in the above equations
 are not independent   
from those with positive one;  they are related by 
an additional phase factor  according to the Wigner-Eckart  
 theorem,     
\begin{equation}    
A^V_{-\lambda}= (-1)^{J_f-J_i-J_V}A^V_{\lambda}    
\end{equation}    
where $J_f$ and $J_i$ are the final nucleon and initial  
 resonance spins, and $J_V$ is the angular momentum of the vector meson.
The angular distributions of the helicity amplitudes in terms of the multipole
transitions have been discussed in Ref. \cite{yang}, the expressions
 here are consistent  with their analysis.

The evaluation of the vector meson helicity amplitudes are 
similar to that of the photon amplitudes. The transition operator for
a resonance decaying into a vector meson and a nucleon is,
\begin{equation}    
H^T_m=\sum_{l} \{i\frac{b^\prime}{2m_q}\vsig_l\cdot({\bf q}
\times\veps_v) +\frac{a}{2\mu_q}{\bf p}^\prime_l  
\cdot\veps_v\}{\hat I}_le^{-i{\bf q\cdot r}_l},     
\end{equation}  
for transverse transitions and
\begin{equation}
H^L_m=\frac{a\mu}{|{\bf q}|}\sum_{l} {\hat I}_le^{-i{\bf q\cdot r}_l}
\end{equation}
for longitudinal transitions. Thus, $H^T_m$ and $H^L_m$ have 
the group structure,
\begin{equation}
H^T_m={\hat I}_3(A L^-_{(3)} +B \sigma^-_{(3)}),
\label{htm}
\end{equation}
and 
\begin{equation}
H^L_m={\hat I}_3 S,
\label{hlm}
\end{equation}
where
\begin{equation}
\label{A}
A=\frac{3 a}{2\sqrt{2}m_q}\langle\psi_f|p^-_3
e^{-i{\bf q}\cdot{\bf r}_3}|\psi_R\rangle,
\end{equation}
\begin{equation}
B=\frac{-3 b^\prime}{2m_q}
|{\bf q}|\langle\psi_f|e^{-i{\bf q}\cdot{\bf r}_3}|\psi_R\rangle,
\end{equation}
\begin{equation}
S=-\frac{3\mu a}{|{\bf q}|}\langle\psi_f|
e^{-i{\bf q}\cdot{\bf r}_3}|\psi_R\rangle.
\end{equation}
where $p^-_3=p_x-ip_y$. 
In Eq.(\ref{htm}), $L^-_{(3)}$ and $\sigma^-_{(3)}$ 
denote orbital and spin flip operators.
The helicity amplitudes $A^V_{\frac 12}$, $A^V_{\frac 32}$
and $S^V_{\frac 12}$ are the matrix elements of Eq.(\ref{htm}) 
and Eq.(\ref{hlm}).
We list the angular momentum and flavor parts of $A^V_{\frac 12}$, 
$A^V_{\frac 32}$ 
and $S^V_{\frac 12}$ for $\omega$ and $\rho$ photoproduction
 in Tables 4-6 in the 
$SU(6)\otimes O(3)$ limit with $A$, $B$ and $S$ in the 
second row to denote the corresponding  spatial integrals, 
which are given in Table 7.

The resonances with $n\ge 3$ are treated as degenerate since there is
little information available about them. Their longitudinal
 transition in the s-channel is given by:
\begin{equation}
h^J_{a\lambda_V=0}=(M^s_3(L)+M^s_2(L))
e^{-\frac{{\bf q}^2+{\bf k}^2}
{6\alpha^2}}
\end{equation}
where 
\begin{eqnarray}
M^s_3(L)&=&g^s_3 \frac{a\mu}{|{\bf q}|}
\{-\frac{i}{2m_q}\vsig\cdot(\veps\times{\bf k}) 
\frac{1}{n!}(\frac{{\bf k}\cdot {\bf q}}{3\alpha^2})^n\nonumber\\
&& +g_v\frac{\omega}{3\alpha^2}{\bf q}\cdot\veps\frac{1}{(n-1)!}
(\frac{{\bf k\cdot 
q}}{3\alpha^2})^{n-1}\},
\label{s-l-3}
\end{eqnarray}
and
\begin{eqnarray}
M^s_2(L)&=&-g^u_2 \frac{a\mu}{|{\bf q}|}
\{g^\prime_a\frac{i}{2m_q}\vsig\cdot(\veps\times{\bf k})
\frac{1}{n!}(\frac{-{\bf k\cdot 
q}}{6\alpha^2})^n\nonumber\\
&&+g^\prime_v\frac{\omega}{6\alpha^2}
{\bf q}\cdot\veps\frac{1}{(n-1)!}(\frac{-{\bf k\cdot 
q}}{6\alpha^2})^{n-1}\}.
\label{s-l-2}
\end{eqnarray}
The $g$-factors in Eq.(\ref{s-l-3}) and (\ref{s-l-2})
have been defined previously, and 
\begin{equation}
g^s_3=\langle N_f|\sum_{j} {\hat I}_j e_j\sigma^z_j|N_i\rangle/g_A=
e_m+g^u_3,
\end{equation}
where $e_m$ is the charge of the outgoing vector meson.

The transverse transition amplitudes at the quark level are:
\begin{equation}
h^J_{a\lambda_V=\pm 1}=(M^s_3(T)+M^s_2(T))
e^{-\frac{{\bf q}^2+{\bf k}^2}{6\alpha^2}}
\end{equation}
where 
\begin{eqnarray}
M^s_3(T)/g^s_3&=&\frac{b^\prime}{4m^2_q}\{g_v({\bf q}\times\veps_v)\cdot(\veps\times{\bf k})
+i\vsig\cdot({\bf q}\times\veps_v)
\times(\veps\times{\bf k})\}\frac{1}{n!}(\frac{{\bf k\cdot 
q}}{3\alpha^2})^n\nonumber\\
&&+\{-\frac{ia}{12m^2_q}\vsig\cdot(\veps\times{\bf k})\veps_v
\cdot{\bf k}
+\frac{ib^\prime\omega}{6m_q\alpha^2}\vsig\cdot({\bf q}\times\veps_v)\veps
\cdot{\bf q}\nonumber\\
&&+g_v\frac{a\omega}{6m_q}\veps_v\cdot\veps\}\frac{1}{(n-1)!}
(\frac{{\bf k}\cdot {\bf q}}{3\alpha^2})^{n-1}\nonumber\\
&&+g_v\frac{a\omega}{18m_q\alpha^2}\veps_v\cdot{\bf k}\veps
\cdot{\bf q} \frac{1}{(n-2)!}
(\frac{{\bf k}\cdot{\bf q}}{3\alpha^2})^{n-2},
\end{eqnarray}
and
\begin{eqnarray}
M^s_2(T)/g^u_2&=&\frac{b^\prime}{4m^2_q}\{g^\prime_v({\bf q}\times\veps_v)
\cdot(\veps\times{\bf 
k})+ig^\prime_a\vsig\cdot({\bf q}\times\veps_v)
\times(\veps\times{\bf k})\}\frac{1}{n!}(\frac{-{\bf 
k\cdot q}}{6\alpha^2})^n\nonumber\\
&&+\{\frac{ia}{24m^2_q}\vsig\cdot(\veps\times{\bf k})\veps_v\cdot{\bf
k}-\frac{ib^\prime\omega}{12m_q\alpha^2}
\vsig\cdot({\bf q}\times\veps_v)\veps\cdot{\bf q}\nonumber\\
&&-g^\prime_v\frac{a\omega}{12m_q}\veps_v\cdot\veps \}
\frac{1}{(n-1)!}(\frac{-{\bf k\cdot 
q}}{6\alpha^2})^{n-1}\nonumber\\
&&+g^\prime_v\frac{a\omega}{72m_q\alpha^2}\veps_v\cdot{\bf k}\veps
\cdot{\bf q} \frac{1}{(n-2)!}(\frac{-{\bf k}\cdot {\bf q}}{6\alpha^2})^{n-2}.
\end{eqnarray}

Qualitatively, we find that the resonances with larger partial waves 
have larger decay widths into the vector meson and nucleon though this is 
not as explicit as in the pseudoscalar case~\cite{pseudoscalar,eta}.
 Thus, we could use the mass
and decay width of the high spin states, such as $G_{17}(2190)$ for $n=3$ 
states and $H_{19}(2220)$ for $n=4$ states in the $\omega$ photoproduction. 
The relation between these operators and the helicity amplitudes
 $h_{a\lambda_V}$ has been given in Table 2 and 3.

\section*{\bf 4. The numerical results}  
Before discussing the details of the $\omega$, $\rho$ and $\phi$ 
productions, it should be pointed out that the nonrelativistic 
wave function in the quark model becomes more inadequate 
as the energy of the system increases.  
A procedure to partly remedy this problem is to introduce 
the Lorentz boost factor in the spatial integrals 
that involve the spatial wavefunctions of nucleons 
and baryon resonances,
\begin{equation}
R(q,k)\to \gamma_q\gamma_k R(q\gamma_q, k\gamma_k),
\end{equation}
where $\gamma_q=\frac{M_f}{E_f}$ and $\gamma_k=\frac{M_i}{E_i}$.  
A similar procedure had been used in the numerical evaluation of
pseudoscalar meson  photoproduction\cite{eta}.  There are two overall
parameters from the quark model formalism; the quark mass $m_q$ and the
parameter $\alpha$ related to the harmonic oscillator strength, and we
adopt the values commonly used in the quark model approach
for these parameters,
\begin{eqnarray}
m_q&=&330\quad\hbox{MeV},\nonumber\\
\alpha &=& 410\quad\hbox{MeV}.
\end{eqnarray}
Now, we turn our attention to the details of 
the $\omega$, $\rho^0$, $\rho^\pm$ and $\phi$ photoproductions.

\subsection*{\bf 4.1. The $\omega$ photoproduction}    
The t-channel exchange of $M^t_{fi}$ in Eq.(\ref{3.1}) would correspond to 
 $\omega$ exchange which is absent since the amplitude is proportional 
to the charge of the outgoing $\omega$ meson.
As discussed in Ref. \cite{FrimanSoyeur}, 
the $\pi^0$ exchange is dominant in the small $t$ region 
over other meson exchanges and largely responsible for the large
diffractive scattering behavior near the threshold.
The Lagrangian for the
$\pi^0$ exchange model has the following form~\cite{FrimanSoyeur},
\begin{equation}\label{3}
L_{\pi NN}=-i g_{\pi NN}\overline\psi \gamma_5(\vtau\cdot\vpi)\psi
\end{equation}
for the $\pi NN$ coupling vertex, and 
\begin{equation}\label{4}
L_{\omega \pi^0 \gamma}=e_N\frac{ g_{\omega\pi\gamma} }{M_\omega}
\epsilon_{\alpha\beta\gamma\delta}\partial^\alpha A^\beta
\partial^\gamma\omega^\delta\pi^0
\end{equation}
for the $\omega\pi\gamma$ coupling vertex, where the $\omega^\delta$ 
and $\pi^0$  represent
 the $\omega$ and $\pi^0$ fields,  the $A^\beta$ denotes the electromagnetic field, 
and  $\epsilon_{\alpha\beta\gamma\delta}$ is the Levi-Civita tensor, and $M_\omega$ 
is the mass of $\omega$ meson. The $ g_{\pi NN}$ and $ g_{\omega\pi\gamma}$ in Eqs. 
(\ref{3}) and (\ref{4}) denote the coupling constants at the two
 vertices, respectively. Therefore, the transition amplitudes of t-channel $\pi^0$ 
exchange have the following expression,
\begin{equation}
M^t_T(\pi^0)= \frac{e_Ng_{\pi NN} g_{\omega\pi\gamma}}{2M_\omega(t-m^2_\pi)}
\{\omega\veps\cdot({\bf q}\times\veps_v)
+\omega_m{\bf k}\cdot(\veps\times\veps_v)\}
\vsig\cdot {\bf A}
e^{-\frac {({\bf q}-{\bf k})^2}{6\alpha_\pi^2}}
\label{t}
\end{equation}
for the transverse transition, and
\begin{equation}
M^t_L(\pi^0)= -\frac{e_Ng_{\pi NN} g_{\omega\pi\gamma}}{2M_\omega(t-m^2_\pi)}
\frac{ M_\omega}{|{\bf q}|}(\veps\times{\bf k})\cdot{\bf q} \vsig\cdot {\bf A}
e^{-\frac {({\bf q}-{\bf k})^2}{6\alpha_\pi^2}}
\label{l}
\end{equation}
for the longitudinal transition, where
$\omega$ in the transition amplitudes
denotes the energy of the photon with momentum ${\bf k}$, and
${\bf A}=-\frac{{\bf q}}{E_f+M_N}+\frac{{\bf k}}{E_i+M_N}$,
and $t=(q-k)^2=M_\omega^2-2k\cdot q$.
The factor $e^{-\frac {({\bf q}-{\bf k})^2}{6\alpha_\pi^2}}$ in Eqs. 
(\ref{t}) and (\ref{l}) is the form factor for both $\pi NN$ and 
$\omega \gamma \pi$ vertices, if we assume
that the wavefunctions for nucleon, $\omega$ and $\pi$ have a Gaussian form.  
The constant 
$\alpha_\pi$ in this form factor is treated as a parameter.   
The coupling constants $ g_{\pi NN}$ and $ g_{\omega\pi\gamma}$
have the values as used in \cite{FrimanSoyeur}. Therefore, 
\begin{eqnarray}
\frac{g^2_{\pi NN}}{4\pi}&=& 14,\nonumber\\
g^2_{\omega\pi\gamma}&=&3.315  \ \ .
\end{eqnarray}
Note that the 
values of $g_{\pi NN}$ and $g_{\omega\pi\gamma}$ were fixed by separate 
experiments 
and, therefore, are not free parameters in Ref.\cite{FrimanSoyeur}.  
Qualitatively, we would
expect that $\alpha_{\pi}$ be smaller than the 
parameter $\alpha=410$ MeV, since it represents 
the combined form factors for both $\pi NN$ and 
$\omega\pi\gamma$ vertices while the parameter
$\alpha$ only corresponds to the form factor for 
the $\pi NN$ or $\omega NN$ vertex alone.
Following the same procedure
as in Section 2, the explicit expressions for the 
operators in terms of the helicity 
amplitudes can be obtained. They are listed in Tables 2 
and 3 for the transverse and 
longitudinal amplitudes, respectively.

As shown in Table 4., the Moorhouse 
selection rule\cite{moor} have eliminated the states 
belonging to $[{\bf 70}, 1^-]_1$ and $[{\bf 70}, 2^+]_2$ representation with 
symmetric spin structure from contributing to the 
$\omega$ photoproduction on the proton
target so that the s-channel states 
$S_{11}(1650), D_{13}(1700), D_{15}(1650)$ are not 
present in our numerical evaluations. 
Of course, configuration mixing will lead to additional
contributions from these resonances  which, 
however, cannot  be determined at present due 
to the poor quality of data.
Only the resonances $P_{13}(1900)$ and
$F_{15}(2000)$, at present classified as 2-star resonances
in the 1996 PDG listings, have masses above the $\omega$ decay threshold, 
and therefore have branching ratio into the $\omega N$ channel. 
We have not performed a rigorous numerical fit to the available 
data because of the poor quality of the data. However,
The numerical results have shown that the resonance $F_{15}(2000)$
plays an very important role in $\omega$ photoproduction.

Fig. 1 shows our calculations for the differential cross section
at the average photon energies of  $E_\gamma=$1.225, 1.45, 1.675 
and 1.915 GeV,  in comparison with the data~\cite{saphir}. The results for
the t-channel $\pi^0$ exchange and contributions from only 
the s- and u-channel processes  are also shown separately. We find that
the remaining paramters in our model are 
\begin{eqnarray}\label{omega}
a & = & -1.7 \nonumber \\
b^\prime & = & 2.5 \nonumber \\
\alpha_{\pi}&=& 300 \quad \hbox{MeV} .
\end{eqnarray}

In order to give a good overall agreement with the data, particularly in
the large $t$ region.  Our results with the $\pi^0$ exchange are consistent
with the findings of Ref. \cite{FrimanSoyeur} though the form factor in
our calculation is different. Fig. 1 clearly demonstrates that
 the t-channel $\pi^0$ exchange is dominant in the small $t$ region,
while the s- and u-channel resonance contributions become more important
as the momentum transfer $t$ increases. 
To test the sensitivity of s-channel resonances to the differential 
cross section, the differential cross section
without the contribution from the resonance $F_{15}(2000)$
at 1.675 GeV (near the threshold of $F_{15}(2000)$ )
is presented in Fig. 1-(c) as well. The results indicate that
the differential cross section data alone are not sufficient to 
determine the presence of this resonance considering the theoretical 
and experimental uncertainties. Since our numerical calculation shows
that the resonance couplings of the
$F_{15}(2000)$ are larger than those of other resonances in this 
mass region, the sensitivity of the differential cross section to
other resonances around 2 GeV is even smaller. 

In contrast to the differential cross section, 
the polarization observables show a much more dramatic dependence 
on the presence of the s-channel resonances. We present
results of four single polarizations at 1.7 GeV in Fig. 2. The absence 
of the resonance $F_{15}(2000)$ leads to a sign change in the 
target polarization, and the variations in the recoil as well 
as the meson polarization observable are very significant as well.
The absence of the resonance 
$P_{13}(1900)$, also shown in Fig. 2,
leads to very significant changes in the recoil 
polarization. Although we 
do not expect our numerical results to give a quantitative
prediction of
polarization observables at the present stage,
since the calculations are limited
to the $SU(6)\otimes O(3)$ symmetry limit that should be broken in 
more realistic quark model wavefunction, our results clearly suggest that the 
polarization observables may be the best place to determine s-channel 
resonance properties.

Our results for the total cross section are shown in Fig. 3, 
in which the contributions from the s- and
u-channel resonances alone are compared to the full calculation. 
Our results indicate
an increasing discrepancy between theory and 
the data~\cite{saphir,ABHMC,olddata}
with increasing energy $E_{\gamma}$,
This discrepancy 
comes mainly from the small angle region where the $\pi^0$ 
exchange alone is not sufficient to describe the
diffractive behavior at higher energies.  One might expect 
that  Pomeron exchange\cite{pomeron,wolf}
plays a more important role in the higher energy region.  
However,  Fig. 1 shows that our results for 
the differential cross section at the large angle region are 
in good agreement with the data, and it 
suggests that contributions from the s- and u- channel resonances
which are the main focus of our study, give
an appropriate description of the reaction mechanism. 

It is interesting to note that the small bump around 1.7 GeV in the total
 cross section
comes from the contributions of the resonance $F_{15}(2000)$.
As discussed above, our calculations
find that the resonance $F_{15}(2000)$
has a strong coupling to the $\omega N$  channel. Thus, 
this resonance is perhaps the best candidate
whose existance as a ``missing" resonance can be 
established through $\omega$ photoproduction.

\subsection*{\bf 4.2. The $\rho^0 $ photoproduction}    
$\rho^0$ meson photoproduction has some similar features 
as $\omega$ photoproduction. The most significant one is that it also has 
strong-forward-peaking diffractive behavior in the 
differential cross section. The t-channel vector meson exchange 
is proportional to the meson charge in (\ref{3.1}),
therefore has no contributions to the transition amplitudes. 
However, since $\rho$ meson is an isovector, its photoproduction
has also shown some different characters from $\omega$ 
photoproduction. There are more s-channel resonances, such as the $\Delta$
resonances, will contribute to the $\rho^0$ productions due to the isospin
couplings.

From the Lagrangian introduced in our quark model approach, 
the amplitudes from s- and u-channel are not sufficient 
to reproduce the diffractive behavior in the small $t$ region. 
Following what we have discussed at the beginning of this 
section, it is reasonable to include an additional t-channel 
meson exchange term in the transition amplitude\cite{collins}. 
As Friman and Soyeur\cite{FrimanSoyeur} has discussed in their 
work that $\sigma$ meson plays dominant role over other 
t-channel processes in the $\rho^0$ photoproduction. As a 
phenomenological approach, $\sigma$ exchange is included to 
give the small $t$ diffractive behavior which can be understood 
as the still sizable contribution from the Pomeron exchange 
from high energies down to the threshold\cite{Freund,Harari}.
In terms of
Regge phenomenology, it corresponds to the nontrivial background 
integral of the Regge trajectory expansion\cite{collins}
which contributes to the diffractive behavior at small
$t$ region. However, we have to mention that, only the resonance
contributions have been taken into account consistently, could 
the proper large $t$ behavior be described. 
The followings are the transverse and longitudinal 
transition amplitudes of $\sigma$ exchange,
\begin{equation}
M^t_T(\sigma)=\frac{ie_N g_{\sigma NN}g_{\rho^0\sigma\gamma}}
{M_\rho(M^2_\rho-2q\cdot k-M^2_\sigma)}
\{-\omega\omega_m\veps\cdot\veps_v+{\bf k}\cdot{\bf q}
\veps\cdot\veps_v-{\bf q}\cdot\veps {\bf k}\cdot\veps_v \}
e^{-\frac{({\bf q}-{\bf k})^2}{6\alpha^2_\sigma}},
\end{equation}
\begin{equation}
M^t_L(\sigma)=\frac{i e_N g_{\sigma NN}g_{\rho^0\sigma\gamma}}
{M^2_\rho(M^2_\rho -2q\cdot k-M^2_\sigma)}
\omega |{\bf q}|{\bf q}\cdot\veps 
e^{-\frac{({\bf q}-{\bf k})^2}{6\alpha^2_\sigma}},
\end{equation}
where $ e^{-\frac{({\bf q}-{\bf k})^2}{6\alpha^2_\sigma}}$ 
is the form factor for t-channel $\sigma$ exchange. 
$\alpha_\sigma $ is treated as the parameter. 
$ g_{\sigma NN}$ and $ g_{\rho^0\sigma\gamma}$ 
are the coupling constants of the vertex
 $\sigma NN$ and $\rho^0\sigma\gamma $, respectively, 
which are determined by the experimental analysis. 
As a phenomenological freedom, the mass of $\sigma$ meson
$M_\sigma$ has the same values as in \cite{FrimanSoyeur}, 
$M_\sigma=500$GeV. $ g_{\sigma NN}$ and 
$ g_{\rho^0\sigma\gamma}$ have the following values,
\begin{eqnarray}
\frac{g^2_{\sigma NN}}{4\pi}&=&8,\nonumber\\
g^2_{\rho^0\sigma\gamma}&=&7.341\quad .
\end{eqnarray}

Numerical investigation indicates that $\pi^0$ exchange
contribution in $\rho^0$ photoproduction is so small that 
can be neglected in the first order approximation. 
With the same values for parameter $a$ and $b^\prime$, 
we fit the differential cross sections, total cross section and
the single spin polarizations in $\gamma p\to \rho^0 p$. 
For $n\le 2$, there are 27 resonances given by the quark model. 
However, Moorhouse selection rule will eliminate those belonging 
to the representation 
$[{\bf 70},1^-]_1,\quad [{\bf 70}, 0^+]_2$ and $[{\bf 70}, 2^+]_2$, 
in which there are $S_{11}(1650),\quad D_{13}(1700)$ 
and $ D_{15}(1675)$ belonging to $[{\bf 70},1^-]_1$,
but those belonging to $[{\bf 70}, 0^+]_2$
and $[{\bf 70}, 2^+]_2$ have not been determined very well in 
experiments.

The experimental data from\cite{saphir} have been 
reproduced in our model 
at $E_\gamma =$1.225, 1.305, 1.4, 1.545, 1.730, 1.925GeV for 
the differential cross sections. We find that the same set of
parameters in the $\omega$ could also be used to describe the $\rho$
productions as well. The additional parameter $\alpha_{\sigma}$ is found
to be 
\begin{equation}
\alpha_{\sigma}=250 MeV.
\end{equation}
The fact that both $\omega$ and $\rho^0$ productions can be described by
the same set of the parameters is by no means trivial. It shows the
advantage of the introduction of the quark degrees of freedom so that the
$\omega$ and $\rho$ can be described by a unified framework, moreover, the
isospin mixings between $\omega$ and $\rho$ is small and they have very
similar masses.  Although the $\sigma$ exchange dominates in the small
$t$ region, s- and
u-channel contributions play obviously important roles in the large $t$ 
region. In fact, it is the s- and u-channel contributions that 
result in the backward peaking behavior which is similar to 
the Compton scattering phenomenon\cite{capstickcompton}. 
In Fig4., the individual results from $\sigma$ exchange and 
s- and u-channel are presented as well. It shows that the cross 
sections are mainly from the diffractive process. Moreover, 
since the number of contributing resonances is large, the 
contribution effects from individual resonance are not 
significant. That is to say, it is quite impossible to derive 
the resonance informations from the differential cross sections.

The polarization results are provided in Fig 5. Much attention
has been paid to the ``missing resonance" $F_{15}(2000)$. 
It shows that the polarization observables are quite sensitive 
to the presence of this state, especially in the recoil 
polarization and meson polarization. Double spin polarization
investigations are in progress in our framework and they should 
be more sensitive to the resonances. 

In Fig 6. all available data\cite{saphir,ABHMC,olddata} for 
$\gamma p\to \rho^0 p$ are present. With the same set of parameters,
 the theoretical 
result give a good description of the experimental data. 
In Fig 6. the dotted line denotes the contributions from s- and u-
channel which shows that the cross section of producing $\rho^0$ 
through resonance channel is quite small in contrast with 
that through diffractive process. This also explains the reason 
that the effects of an individual resonance is not significant 
in the differential cross sections.

\subsection*{\bf 4.3. $\rho^\pm $ photoproduction}    
In the charge meson productions,  $\gamma n\to \rho^- p$ and
$\gamma p\to \rho^+ n$, the  three channels s-, u- and t- have
contributions to the 
transition amplitudes. The charge exchange process has 
eliminated contributions from such diffractive behaviors 
as in the $\omega$ and $\rho^0$ photoproductions while 
the t-channel charged vector meson exchange and the Seagull
term will account for the small forward peaking shapes 
of the differential cross sections, and this is also required 
by the duality hypothesis when we have taken 
the contributions from all the s- and u-channel processes 
into account.  Therefore, the charged meson productions provides an
important test of this approach, as every term in these two reactions are
generated by the effective Lagrangian, and there is no additional free
parameters.

In $\gamma p\to \rho^+ n$, for the Moorhouse selection rule 
at the photon interaction vertex, those resonances belonging 
to $[{\bf 70},1^-]_1,\quad [{\bf 70}, 0^+]_2$ 
and $[{\bf 70}, 2^+]_2$ are eliminated from contributing to 
the amplitudes. But in $\gamma n\to \rho^- p$, without 
the constraint from Moorhouse selection rule, more resonances 
have contributions to the amplitudes. 

As we have shown in the $\omega$ and $\rho^0$ photoproductions, 
the same set of parameters $a$ and $b^\prime$ gives an overall 
agreement with the available data, the challenge is that 
whether we can reproduce the data\cite{benz} in the $\rho^\pm$
channels with the same parameters,
because they possess the same isospin symmetry. This should be 
one crucial test for the model. As expected, 
the data in the reaction $\gamma n\to \rho^- p$ are in 
very good agreement
with the quark model predictions, indicating that the quark model
wave functions appear to provide the correct relative strengths 
and phases among 
the terms in the s-, u- and t-channels. In Fig. 7(a)., 
the experimental data for $\gamma n\to \rho^- p$
from Ref.\cite{benz} are presented at $E_\gamma=1.85$GeV, 
which is the average of the measurement realm. In Fig. 7(b).,
we present the prediction of the differential cross section 
in $\gamma p\to \rho^+ n$, and it shows a similar behavior 
as in $\gamma n\to \rho^- p$. 

In Fig. 8., the total cross sections for $\rho^\pm$ \cite{benz}
are presented. In Fig. 8(a), the cross section of $\rho^-$ 
photoproduction
with $0<|t|<1.1\hbox{GeV}^2$ is well reproduced, 
along which the total cross section is also consistent with 
the experimental estimation\cite{hilpert}. The prediction 
of the total cross section of $\rho^+$ photoproduction is 
given in Fig. 8(b).

While the shapes and magnitudes of the differential cross sections are
well reproduced within our approach we find little sensitivity to
individual resonances.  For example,
in the energy region of $E_{\gamma}\sim 1.7$GeV, removing the
$F_{15}(2000)$ state - one of the ``missing" candidates -
changes the cross section very little, indicating
the differential cross section may not be the
 ideal experimental observable to study the structure
of the baryon resonances.
In contrast to the cross sections, the polarization observables
show a more dramatic dependence on the presence of the s-channel
resonances. To illustrate their effects, as an example,
the target polarizations for $\rho^-$ and $\rho^+$ 
production with and without the contribution from
the $F_{15}(2000)$ resonance are shown in Fig. 9.  
We do not expect the quark model in the 
$SU(6)\otimes O(3)$ limit to provide a good description of these observables.
However, it demonstrates the sensitivity of  these observables
to the presence of s-channel resonances.
This shows that  polarization observables are essential
in analyzing the role of s-channel resonances.

\subsection*{\bf 4.4. The $\phi$ photoproduction}

Because the isospin of the $\phi$ is the same as that of the
$\omega$, the formalism for the s- and u- channel contributions
to the $\phi$ productions should be the same as that of the $\omega$ 
productions except the different threshold energies. The major 
difference between the $\omega$ and the $\phi$ productions is the 
mechanism of generating $u$ ($d$) and $\overline u$ ($\overline d$) 
quarks for the $\omega$ productions and the $s$ and $\overline s$ quarks
for the $\phi$ production that are suppressed by the OZI rule. Such a
difference will be reflected in the difference of
the coupling constants between
$\omega$ and $\phi$ productions, of which
the coupling constant for the $\phi NN$
vertex is expected to be much smaller than that for the $\omega NN$ vertex.
Thus, we shall concentrate on the non-diffractive effects generated 
from the effective Lagrangian in the s- and u-channel reaction. 
Since we have not included the Pomeron exchange term to give the strong 
diffractive behavior in the small $t$  region, our results thus could
only be regarded as an estimation.

The numerical results with the following two sets of parameters 
are shown in comparison with the data\cite{phidata} in Fig.10.
They represent the $\phi qq$ coupling constants in the non-diffractive
processes of the $\phi$ production,
\begin{eqnarray}\label{phi1}
a&=&-0.35, \nonumber\\
b^\prime &=& 0.7\quad ,
\end{eqnarray}
and 
\begin{eqnarray}\label{phi2}
a&=&-0.6, \nonumber\\
b^\prime &=& 1.2 \quad .
\end{eqnarray}
It should be note that the same value of $\alpha$ as
that in the $\omega$ production
has been used in the evaluations at present stage.
Our results show that the differential cross sections are very
senstive to the parameters $a$ and $b^\prime$ in the large $t$
region, where the Pomeron exchange is expected to be small. Thus,
the data in the large $t$ region could provide an important 
constraint to the $\phi NN$ coupling constants. Moreover, the
coupling constants in Eqs. (\ref{phi1}) and (\ref{phi2}) are indeed
significantly smaller than those for the $\omega$ photoproductions
in Eq.(\ref{omega}). These results are also consistent with those
obtained in the QHD approach\cite{williams}.

\section*{\bf 5. Discussion and conclusion}    

In this paper we have developed the framework and formalism
for the description of the vector meson photoproductions in the 
constituent quark model. Consequently, the application 
of this approach to the
$\omega$ and $\rho$ meson photoproductions have 
produced very encouraging results. 
The use of an effective Lagrangian allows the gauge invariance to be
satisfied straightforwardly.

The advantage of using the quark model approach is that
the number of free parameters is greatly reduced in comparison
with hadronic models which introduce each resonance as a new
independent field with unknown coupling constants.
In our approach, only three parameters appear
in the $SU(6)\otimes O(3)$ symmetry limit, 
the coupling constants $a$ and $b$ (or $b^\prime$) which determine the
coupling strengths of the vector meson to the quark,
and the harmonic oscillator strength $\alpha$. 

With $\pi^0$ and $\sigma$ exchange taken into account, 
an overall description of the $\omega$,
$\rho^0$, $\rho^+$ and $\rho^-$ photoproduction 
with the same set of parameters has been obtained 
in this framework. It shows that intermediate 
resonance contributions have played 
important roles in the $\omega$ and $\rho$ 
meson photoproductions especially in large $t$ regions. 
This shows that our effective Lagrangian approach in the quark model 
has provided an ideal framework to investigate the reaction 
mechanism and the underlying quark structure of the baryon 
resonances. The crucial role played by the polarization 
observables in determining
the s-channel resonance properties is demonstrated.
Data on these observables, expected from TJNAF in the near future,
should therefore provide new insights into the
structure of the resonance $F_{15}(2000)$ 
as well as other ``missing" resonances.

The introduction of t-channel $\pi^0$ and $\sigma$ exchange 
in the neutral productions can be quanlitively understood 
in the picture of the Regge phenomenology\cite{collins} 
or the diffraction duality 
picture of Freund\cite{Freund} and Harari\cite{Harari}.
In such a picture, 
the large difference of the cross section 
between $\rho^0$ and $\rho^\pm$ is due to such a background 
amplitude which originates from a sizable contribution 
of the Pomeron singularity in $\rho^0$ photoproduction 
from high energies down to the threshold. 
Our numerical investigation has really shown the case, 
therefore, to some extent, suggests that the 
duality hypothesis\cite{duality}
constrains also the vector meson photoproductions.

In the reaction of $\phi$ photoproduction, the sizable non-diffractive 
contributions can be phenomenologically interpreted as the 
s- and u- channel contributions generated from the effective Lagiangian. 
Further studies that includes the pomeron exchange in this approach
will be persued later. 

One significant approximation inherent in the presented approach
is the treatment of the vector mesons as point particles, thus,
the effects due to the finite size 
of the vector mesons that were important in the $^3P_0$ model
are neglected here.
A possible way that may partly compensate this problem 
is to adjust the parameter $\alpha^2$,  the harmonic
oscillator strength. 
In general, the question of how to include the finite 
size of vector mesons 
while maintaining the gauge invariance is 
very complicated and has not yet been resolved.
Moreover, as configuration mixing effects for the 
resonances in the second and third resonance 
region are known to be very important, more precise 
quantitative agreement with 
the data cannot be derived from the current form.
But such effects could be investigated 
in our approach by inserting a 
mixing parameter $C_R$ in front of the transition 
amplitudes for the  s-channel 
resonances, as has been investigated in Ref.~\cite{eta}. 

The fact that the $\omega$ and $\rho$ productions 
can be described by the same set of the parameters shows the
successes of the quark model appraoch. Thus,  the model
presented here could provide a systematic method to investigate 
the resonance behavior in the vector meson photoproductions
for the first time,
which will help us  to identify the ``missing resonances".

The authors acknowledge useful discussions 
with B. Saghai, F. Tabakin and P. Cole.
Helpful discussions with F. J. Klein regarding 
the data are acknowledged gratefully. 
Zhenping Li acknowledges the hospitality of the 
Saclay Nuclear Physics Laboratory. Q. Zhao acknowledges helpful 
discussions with Yang Ze-sen and Hongan Peng. 
This work is supported in part by Chinese Education 
Commission and Peking University, and the US-DOE 
grant no. DE-FG02-95-ER40907.

\section*{\bf Appendix}    

The matrix element for the nucleon pole term of transverse 
excitations in the s-channel is,
\begin{eqnarray}
M^s_N(T)&=&-\frac{M_N
e^{-\frac{{\bf q}^2+{\bf k}^2}{6\alpha^2}}}{P_N\cdot k}
\{g^t_v\frac{\omega ae_N}{E_f+M_f}
\veps_v\cdot\veps -g_A\mu_N\frac{b^\prime}{2m_q}
[(\veps_v\times {\bf q})\cdot(\veps
\times {\bf k})\nonumber\\
&&+i\vsig\cdot(\veps_v\times {\bf q})\times(\veps\times {\bf k})]\},
\end{eqnarray}
while the one for the u-channel is,
\begin{eqnarray}
M^u_N(T) & = & -\frac{M_fe^{-\frac{{\bf q}^2
+{\bf k}^2}{6\alpha^2}}}{P_f\cdot k}
\{ g^t_v\frac{\omega ae_f}{E_N+M_N}\veps\cdot\veps_v
+g_A\mu_f\frac{b^\prime}{2m_q}
\big [(\veps\times {\bf k})\cdot(\veps_v\times {\bf q})
  \nonumber  \\ 
& &  + i\vsig\cdot((\veps\times {\bf k})\times(\veps_v\times {\bf q}))
\big ]\}\nonumber\\
&&+ \frac{ e_f e^{-\frac{{\bf q}^2+{\bf k}^2}{6\alpha^2}}}
{P_f\cdot k}\{\frac{-g^t_v a}{E_N+M_N}{\bf q}\cdot\veps{\bf k}\cdot\veps_v
 +ig_A\frac{b^\prime}{2m_q}\vsig\cdot(\veps_v\times {\bf q})
{\bf q}\cdot\veps\}. 
\end{eqnarray}

The matrix element for the nucleon pole term of the longitudinal 
excitations in the s-channel is,
\begin{equation}
M^s_N(L)=-g^t_v\frac{i\mu a }{|{\bf q}|}
\frac{(w+M_N)}{2 P_N\cdot k}\mu_N\vsig\cdot(\veps\times{\bf k})
e^{-\frac{{\bf q}^2
+{\bf k}^2}{6\alpha^2}},
\end{equation}
while the one for the u-channel is, 
\begin{equation}
M^u_N(L)=g^t_v\frac{\mu a }{|{\bf q}|}\frac{1}{P_f\cdot k}\{ -e_f{\bf q}
\cdot \veps+i\mu_f \frac{(w+M_f)}{2w} \vsig\cdot(\veps\times{\bf k})\}
e^{-\frac{{\bf q}^2+{\bf k}^2}{6\alpha^2}},
\end{equation}
where $w=E_i+\omega=E_f+\omega_m$ is the c.m. energy and 
the $g$-factor $g^t_v$ has been given in (\ref{gtv}).

The t-channel matrix element for the transverse transition is,
\begin{eqnarray}
M^t(T)&=&-\frac{a e_m}{q\cdot k}
\{-g^t_v[\omega_m+(\frac{{\bf q}}{E_f+M_f}
+\frac{{\bf k}}{E_N+M_N})\cdot{\bf q}]\veps\cdot\veps_v
\nonumber\\
&&+g_A\frac{i}{2m_q}\vsig\cdot({\bf k}
\times{\bf q})\veps\cdot\veps_v\nonumber\\
&&-g^t_v(\frac{1}{E_f+M_f}+\frac{1}{E_N+M_N})
{\bf q}\cdot\veps{\bf k}\cdot\veps_v \nonumber\\
&&+g_A\frac{i}{2m_q}\vsig\cdot(({\bf k}-{\bf q})
\times\veps_v){\bf q}\cdot\veps\nonumber\\
&&+g_A\frac{i}{2m_q}\vsig\cdot(({\bf k}-{\bf q})
\times\veps){\bf k}\cdot\veps_v
 \}e^{-\frac{({\bf k}-{\bf q})^2}{6\alpha^2}},
\end{eqnarray}
and for the longitudinal transition is,
\begin{equation}
M^t(L)=-\frac{\mu}{|{\bf q}|}
\frac{ae_m}{q\cdot k}\{g^t_v(1-\frac{\omega}{E_f+M_f})
{\bf q}\cdot\veps
+g_A\frac{i\omega}{2m_q}\vsig\cdot(({\bf k}-{\bf q})
\times\veps) \}e^{-\frac{({\bf k}-{\bf q})^2}{6\alpha^2}}.
\end{equation}

\newpage
\begin{table}
\caption{The $g$-factors in the u-channel amplitudes
in Eqs.(\ref{amp3})
and (\ref{amp2}) for different production processes.}
\protect\label{tab:(1)}
\begin{center}
\begin{tabular}{lccccccccc}
\hline
Reactions & & $g^u_3$ & $g^u_2$ & $g_v$ & $g_v^\prime$& $g_a^\prime$ 
& $g_A$ & $g_S$&$g^t_v$\\[1ex]\hline
$\gamma p\to \omega p$ & & 1 & 0 & 1 & 0 & 0&  1 & 0 &3\\[1ex]
$\gamma n\to \omega n$ & &  -$\frac 23$  & $\frac 23$ & 0 
& -1&  0&  1 & 0 &3\\[1ex]
$\gamma p\to \rho^+ n$ & & -$\frac 13$  & $\frac 13$ & $\frac 35$ 
&$\frac 15$ & $\frac 95$ & $\frac 53$ & -$\frac{2\mu_n}5$ &1\\[1ex]
$\gamma n\to \rho^- p$ & & $\frac 23$ & $\frac 13$ & $\frac 35$ 
&$\frac 15$ &- $\frac 95$  &  - $\frac 53$ & -$\frac{4\mu_p}{15}$&1\\[1ex]
$\gamma p\to \rho^0 p$ & & $\frac 7{15}$ & $\frac 8{15}$ 
&$\frac {15}7$ &  2 & 0 &  $\frac 5{3\sqrt{2}}$ & $\frac{8\mu_p}{15}$
&$-\frac{1}{\sqrt 2}$\\[1ex]
$\gamma n\to \rho^0 n$ & & -$\frac 2{15}$ & $\frac 2{15}$ & 6 &  -7 
& 0 & -$\frac 5{3\sqrt{2}}$ & $\frac{4\mu_n}5$&$\frac{1}{\sqrt 2}$
\\[1ex]
$\gamma p\to K^{*+}\Lambda$ & & -$\frac 13$ &  $\frac 13$ & 1 & 1 
&1& $\sqrt{\frac 32}$ & -$\frac {\mu_{\Lambda}}3$&$\sqrt{\frac 32}$\\[1ex]
$\gamma n\to K^{*0}\Lambda$ & & -$\frac 13$ & $\frac 13$ & 1 & -1 
&-1&  $\sqrt{\frac 32}$ & $\frac {\mu_{\Lambda}} 3$&$\sqrt{\frac 32}$\\[1ex]
$\gamma p \to K^{*+}\Sigma^0$ & & -$\frac 13$ & $\frac 13$ & -3 & -7
& 9 & -$\frac 1{3\sqrt{2}}$ & $\mu_{\Sigma^0}$&$\frac{1}{\sqrt 2}$\\[1ex]
$\gamma n\to K^{*0}\Sigma^0$ & & -$\frac 13$ & $\frac 13$ & -3 & 11 
&-9& $\frac 1{3\sqrt{2}}$ &  $\mu_{\Sigma^0}$&$-\frac{1}{\sqrt 2}$\\[1ex]
$\gamma p\to K^{*0}\Sigma^+$ & & -$\frac 13$ & $\frac 43$ & -3 & 2 
&0 & $\frac 13$ &  $\frac {2\mu_{\Sigma^+}}{3}$ &0\\[1ex]
$\gamma n\to K^{*+}\Sigma^-$ & & -$\frac 13$& -$\frac 23$ & -3 & 2 
&0&  -$\frac 13$ &  0&1\\[1ex]
$\gamma p\to \phi p$ & & 1 & 0 & 1 & 0 & 0&  1 & 0 &3\\[1ex]
$\gamma n\to \phi n$ & &  -$\frac 23$  & $\frac 23$ & 0 
& -1&  0&  1 & 0 &3\\[1ex]
\hline
\end{tabular}
\end{center}
\end{table}    

\begin{table}
\caption{ The operators in the longitudinal excitations expressed 
in terms of the helicity amplitudes.
${\hat{\bf k}}$ and ${\hat{\bf q}}$ 
are the unit vectors of ${\bf k}$ and  ${\bf q}$, respectively.
The $d$ functions depend on
the rotation angle $\theta$ between {\bf k} and {\bf q}. 
All other components of $H_{a\lambda_V}$ are zero. $\lambda_f =\pm\frac 12$ 
denotes the helicity of the final state nucleon. }
\protect\label{tab:(2)}
\begin{center}
\begin{tabular}{lcll}
\hline
Operators&
&$H_{10}(\lambda_f=\frac 12)$&$H_{20}(\lambda_f=\frac 12)$\\
&&$H_{30}(\lambda_f=-\frac 12)$&$H_{40}(\lambda_f=-\frac 12)$\\[1ex]\hline
${\hat{\bf q}}\cdot \veps$
&&$d^1_{10}d^{\frac 12}_{-\frac 12\lambda_2 }$
&$d^1_{10} d^{\frac 12}_{\frac 12\lambda_2 } $
\\[1ex]\hline
$\vsig\cdot(\veps\times{\hat{\bf k}})$
&&$i\sqrt{2} d^{\frac 12}_{\frac 12\lambda_2 }$
&0\\[1ex]\hline
$\vsig\cdot(\veps\times{\hat{\bf q}})$
&&$ id^1_{10} d^{\frac 12}_{-\frac 12\lambda_2 }$
&$-id^1_{10} d^{\frac 12}_{\frac 12\lambda_2 }$\\
&&$+i\sqrt{2}d^1_{00}d^{\frac 12}_{\frac 12\lambda_2 } $& \\[1ex]\hline 
$(\veps\times{\hat{\bf k}})\cdot{\hat{\bf q}}\vsig\cdot{\hat{\bf q}}$&
&$i\sqrt{2}d^1_{10}  d^1_{10} d^{\frac 12}_{\frac 12\lambda_f} $
&$ i\sqrt{2}d^1_{10}  d^1_{-10} d^{\frac 12}_{-\frac 12\lambda_f } $\\[1ex]
&&$ -id^1_{10}  d^1_{00} d^{\frac 12}_{-\frac 12\lambda_f } $
&$ +id^1_{10}  d^1_{00} d^{\frac 12}_{\frac 12\lambda_f } $\\[1ex]\hline
$(\veps\times{\hat{\bf k}})\cdot{\hat{\bf q}}\vsig\cdot{\hat{\bf k}}$&
&$-id^1_{10}d^{\frac 12}_{-\frac 12 \lambda_f }$
&$id^1_{10}d^{\frac 12}_{\frac 12 \lambda_f }$\\[1ex]\hline
\end{tabular}    
\end{center}
\end{table}    

\begin{table}
\caption{ The operators in the transverse excitations expressed 
in terms of the helicity amplitudes.
${\hat{\bf k}}$ and ${\hat{\bf q}}$
 are the unit vectors of ${\bf k}$ and ${\bf 
q}$, respectively. The $d$ functions depend on the rotation angle 
$\theta$ between {\bf k} and {\bf q}.
$\lambda_V=\pm 1$ denotes the helicity of $\omega$ meson, 
and $\lambda_f=\pm\frac 12$ denotes the helicity of the final nucleon. }
\protect\label{tab:(3)}
\begin{center}
\begin{tabular}{lcll}    
\hline
Operators&
&$H_{1\lambda_V}(\lambda_f=\frac 12) $
&$H_{2\lambda_V}(\lambda_f=\frac 12) $\\
&&$H_{3\lambda_V}(\lambda_f=-\frac 12) $
&$H_{4\lambda_V}(\lambda_f=-\frac 12)$\\[1ex] \hline

$(\veps\times{\hat{\bf k}})\cdot(\veps_v\times{\hat{\bf q}})$
&&$-\lambda_V d^1_{1\lambda_V} d^{\frac 12}_{-\frac 12\lambda_2}$
&$-\lambda_V d^1_{1\lambda_V} d^{\frac 12}_{\frac 12\lambda_2 }$
\\[1ex]\hline

$\vsig\cdot{\hat{\bf Z}}$
&&$-i\lambda_V d^1_{11} d^{\frac 12}_{-\frac 12\lambda_2 }$
&$i\lambda_V d^1_{11} d^{\frac 12}_{\frac 12\lambda_2 }$\\
&&$-i\sqrt{2}\lambda_V d^1_{0\lambda_V} d^{\frac 12}_{\frac 12\lambda_2 }$
&\\[1ex]\hline

$ \vsig \cdot(\veps\times{\hat{\bf k}}){\hat{\bf k}}\cdot\veps_v $
&&$i\sqrt{2}d^1_{0\lambda_V} d^{\frac 12}_{\frac 12\lambda_2 }$&0
\\[1ex]\hline

$ \vsig \cdot(\veps_v\times{\hat{\bf q}}){\hat{\bf q}}\cdot\veps $
&&$i\sqrt{2} \lambda_Vd^1_{10}d^1_{1\lambda_V} d^{\frac 12}_{\frac 12\lambda_2 }$
&$i\sqrt{2} \lambda_Vd^1_{10}d^1_{-1\lambda_V} d^{\frac 12}_{-\frac 12\lambda_2 }$\\
&&$-i \lambda_Vd^1_{10}d^1_{0\lambda_V} d^{\frac 12}_{-\frac 12\lambda_2 }$
&$+i \lambda_Vd^1_{10}d^1_{0\lambda_V} d^{\frac 12}_{\frac 12\lambda_2 }$
\\[1ex]\hline

$\veps_v\cdot\veps$
&&$ d^1_{1\lambda_V} d^{\frac 12}_{-\frac 12\lambda_2 }$
&$d^1_{1\lambda_V} d^{\frac 12}_{\frac 12\lambda_2 }$\\[1ex]\hline

${\hat{\bf q}}\cdot\veps {\hat{\bf k}}\cdot\veps_v $
&&$d^1_{0\lambda_V}d^1_{01} d^{\frac 12}_{-\frac 12\lambda_2 }$
&$d^1_{0\lambda_V}d^1_{01} d^{\frac 12}_{\frac 12\lambda_2 }$
\\[1ex]\hline

$ \vsig \cdot(\veps\times{\hat{\bf q}}){\hat{\bf k}}\cdot\veps_v $
&&$i\sqrt{2}d^1_{00}d^1_{0\lambda_V} d^{\frac 12}_{\frac 12\lambda_2 }$&\\
&&$+id^1_{10}d^1_{0\lambda_V} d^{\frac 12}_{\frac 12\lambda_2 }$
&$-id^1_{10}d^1_{0\lambda_V} d^{\frac 12}_{\frac 12\lambda_2 }$
\\[1ex]\hline

$ \vsig \cdot(\veps_v\times{\hat{\bf k}}){\hat{\bf q}}\cdot\veps $
&&$-i\sqrt 2d^1_{10}d^1_{-1\lambda_V} d^{\frac 12}_{\frac 12\lambda_2 }$
&$-i\sqrt 2d^1_{10}d^1_{1\lambda_V} d^{\frac 12}_{-\frac 12\lambda_2 }$
\\[1ex]\hline

$\vsig\cdot(\veps_v\times\veps)$
&&$-i\sqrt 2d^1_{0\lambda_V} d^{\frac 12}_{\frac 12\lambda_2 }$&\\
&&$-id^1_{1\lambda_V} d^{\frac 12}_{-\frac 12\lambda_2 }$
&$id^1_{1\lambda_V} d^{\frac 12}_{\frac 12\lambda_2 }$
\\[1ex]\hline

$\vsig \cdot({\hat{\bf k}}\times{\hat{\bf q}})\veps\cdot\veps_v $
&&$i\sqrt 2d^1_{-10}d^1_{1\lambda_V} d^{\frac 12}_{\frac 12\lambda_2 }$
&$i\sqrt 2d^1_{10}d^1_{1\lambda_V} d^{\frac 12}_{-\frac 12\lambda_2 }$
\\[1ex]\hline 

$ (\veps \times\veps_v) \cdot {\hat{\bf q}}\vsig\cdot{\hat{\bf q}}$&
&$i\sqrt{2}\lambda_V d^1_{1\lambda_V}d^1_{10}d^{\frac 12}_{\frac 12\lambda_f}$
&$i\sqrt{2}\lambda_V d^1_{1\lambda_V}d^1_{-10}d^{\frac 12}_{-\frac 12\lambda_f }$\\[1ex]
&&$-i\lambda_V d^1_{1\lambda_V}d^1_{00}d^{\frac 12}_{-\frac 12\lambda_f }$
&$+i\lambda_V d^1_{1\lambda_V}d^1_{00}d^{\frac 12}_{\frac 12\lambda_f }$\\[1ex]\hline
$ (\veps \times \veps_v) \cdot {\hat{\bf q}}\vsig\cdot{\hat{\bf k}}$&
&$-i\lambda_V d^1_{1\lambda_V}d^{\frac 12}_{-\frac 12\lambda_f }$
&$i\lambda_V d^1_{1\lambda_V}d^{\frac 12}_{\frac 12\lambda_f }$\\[1ex]\hline

$ (\veps\times\veps_v) \cdot {\hat{\bf k}}\vsig\cdot{\hat{\bf q}}$&
&$i\sqrt{2} d^1_{1\lambda_V}d^1_{10}d^{\frac 12}_{\frac 12\lambda_f }$
&$i\sqrt{2} d^1_{1\lambda_V}d^1_{-10}d^{\frac 12}_{-\frac 12\lambda_f }$\\[1ex]
&&$-i  d^1_{1\lambda_V}d^1_{00}d^{\frac 12}_{-\frac 12\lambda_f }$
&$+i  d^1_{1\lambda_V}d^1_{00}d^{\frac 12}_{\frac 12\lambda_f }$\\[1ex]\hline
$ (\veps\times\veps_v)\cdot {\hat{\bf k}}\vsig\cdot{\hat{\bf k}}$&
&$-i d^1_{1\lambda_V}d^{\frac 12}_{-\frac 12\lambda_f }$
&$i d^1_{1\lambda_V}d^{\frac 12}_{\frac 12\lambda_f }$\\[1ex]\hline 
\end{tabular}
\end{center}
\end{table}

\begin{table}
\caption{ The agular momentum and flavor parts of the 
helicity amplitudes for $\gamma p\to \omega p (\phi p)$,
or $\gamma n\to \omega n (\phi n)$, in the $SU(6)\otimes O(3)$ symmetry limit.
They are the coefficients of the spatial integrals $A$,
$B$ and $S$ in Eq.(\ref{htm}) and (\ref{hlm}).
The analytic expressions for $A$, $B$ and $S$ in
Table 4-6 are given in Table 7. The ``*" in Table 4-6 denotes
those states decoupling in spin and flavor space, thus,
their amplitudes are zero.}
\protect\label{tab:(4)}
\begin{center}
\begin{tabular}{lcccccccccc}
\hline    
States & & $S_{1/2}$ & & &$A_{1/2}$ & & & & $A_{3/2}$  &\\[1ex] \hline     
 & & S & & A & & B & & A & & B\\[1ex] \hline    
$N(^2P_M)_{\frac 12^-}$&&0&&0&&$-\frac 2{3\sqrt{6}}$&&0&&0\\[1ex]    
$ N(^2P_M)_{\frac 32^-}$&&0&&0&&$\frac 2{3\sqrt{3}}$&&0&&0\\[1ex]    
$ N(^4P_M)_{\frac 12^-}$&&0&&0&&$-\frac 1{3\sqrt{6}}$&&*&&*\\[1ex]    
$ N(^4P_M)_{\frac 32^-}$&&0&&0&&$\frac 1{3\sqrt{30}}$&&0&&$-\frac1{\sqrt{15}}$\\[1ex]    
$ N(^4P_M)_{\frac 52^-}$&&0&&0&&$\frac 1{\sqrt{30}}$&&0&&$\frac1{\sqrt{10}}$\\[1ex]    
$N(^2D_S)_{\frac 32^+}$&&$-\sqrt{\frac 25}$&&$\sqrt{\frac 35}
$&&$-\frac 13 \sqrt{\frac 25}$&&$-\frac1{\sqrt{5}}$&&*\\[1ex]    
$N(^2D_S)_{\frac 52^+}$&&$\sqrt{\frac 35}$&&$\sqrt{\frac 25}
$&&$\frac 13 \sqrt{\frac 35}$&&$\frac2{\sqrt{5}}$&&*\\[1ex]    
$N(^2S^\prime_S)_{\frac 12^+}$&&1&&*&&$\frac 13$&&*&&*\\[1ex]    
$N(^2S_M)_{\frac 12^+}$&&0&&*&&$\frac 2{3\sqrt{2}}$&&*&&*\\[1ex]    
$N(^4S_M)_{\frac 32^+}$&&0&&*&&$\frac 1{3\sqrt{2}}$&&*&&$\frac 1{\sqrt{6}}$\\[1ex]
$N(^2D_M)_{\frac 32^+}$&&0&&0&&$-\frac 2{3\sqrt{5}}$&&0&&*\\[1ex]    
$N(^2D_M)_{\frac 52^+}$&&0&&0&&$\sqrt{\frac 2{15}}$&&0&&*\\[1ex]    
$N(^4D_M)_{\frac 12^+}$&&0&&0&&$\frac 1{3\sqrt{10}}$&&*&&*\\[1ex]    
$N(^4D_M)_{\frac 32^+}$&&0&&0&&$-\frac 1{3\sqrt{10}}$&&0&&$\frac1{\sqrt{30}}$\\[1ex]    
$N(^4D_M)_{\frac 52^+}$&&0&&0&&$-\frac 1{\sqrt{210}}$&&0&&
$-\sqrt{\frac 3{35}}$\\[1ex]
$N(^4D_M)_{\frac 72^+}$&&0&&0&&$\frac 1{\sqrt{35}}$&&0&&$\frac 1{\sqrt{21}}$\\[1ex]
\hline    
\end{tabular}
\end{center}
\end{table}    

\begin{table}
\caption{ The angular momentum and flavor parts of the 
helicity amplitudes for $\gamma p\to \rho^0 p$ in the
$SU(6)\otimes O(3)$ symmetry limit, while those for
$\gamma n\to\rho^0 n$ are given by $A(\gamma n\to \rho^0 n)=
(-1)^{I+1/2}A(\gamma p\to \rho^0 p)$, where $I$ is the
isospin of the resonances.}
\protect\label{tab:(5)}
\begin{center}
\begin{tabular}{lcccccccccc}
\hline   
States & & $S_{1/2}$ & & &$A_{1/2}$ & & & & $A_{3/2}$  
&\\[1ex] \hline     
 & & S & & A & & B & & A & & B\\[1ex]\hline    
$N(^2P_M)_{\frac 12^-}$&&$\frac 1{3\sqrt{3}}$&&$-\frac{\sqrt 2}{3\sqrt3}$&&$\frac 2{9\sqrt{3}}$&&*&&*\\[1ex]    
$ N(^2P_M)_{\frac 32^-}$&&$-\frac 13\sqrt{\frac 23}$&&$-\frac 1{3\sqrt{3}}$&&$-\frac 29\sqrt{\frac 23}$&&$-\frac 13$&&*\\[1ex]    
$ N(^4P_M)_{\frac 12^-}$&&0&&0&&$-\frac 1{18\sqrt{3}}$&&*&&*\\[1ex]    
$ N(^4P_M)_{\frac 32^-}$&&0&&0&&$\frac 1{18\sqrt{15}}$    
&&0&&$\frac 1{6\sqrt5}$\\[1ex]    
$ N(^4P_M)_{\frac 52^-}$&&0&&0&&$\frac 1{6\sqrt{15}}$&&0&&$
\frac 1{\sqrt{30}}$\\[1ex]    
$\Delta(^2P_M)_{\frac 12^-}$&&$-\frac 1{3\sqrt3}$&&$\frac 13
\sqrt{\frac 23}$&&$\frac 1{9\sqrt3}$&&*&&*\\[1ex]    
$\Delta(^2P_M)_{\frac 32^-}$&&$\frac 13\sqrt{\frac 23}$&&$\frac 1{3\sqrt3}$&&$-\frac 19\sqrt{\frac 23}$&&$\frac 13$&&*\\[1ex]    
$N(^2D_S)_{\frac 32^+}$&&$\frac 1{3\sqrt5}$&&$-\frac 1{\sqrt{30}}$
&&$\frac{\sqrt{5}}{9}$&&$\frac1{3\sqrt{10}}$&&*\\[1ex]    
$N(^2D_S)_{\frac 52^+}$&&$-\frac 1{\sqrt{30}}$&&$-\frac 1{3\sqrt5}$
&&$-\frac 59\sqrt{\frac 3{10}}$&&$-\frac2{3\sqrt{10}}$&&*\\[1ex]    
$\Delta(^4D_S)_{\frac 12^+}$&&0&&0&&$\frac 2{9\sqrt5}$&&*&&*\\[1ex]    
$\Delta(^4D_S)_{\frac 32^+}$&&0&&0&&$-\frac 2{9\sqrt5}$&&0&&$\frac 2{3\sqrt{15}}$\\[1ex]    
$\Delta(^4D_S)_{\frac 52^+}$&&0&&0&&$-\frac 29\sqrt{\frac 3{35}}$
&&0&&$-\frac 23\sqrt{\frac 6{35}}$\\[1ex]    
$\Delta(^4D_S)_{\frac 72^+}$&&0&&0&&$\frac 23\sqrt{\frac2{35}}$
&&0&&$\frac 4{3\sqrt{42}}$\\[1ex]    
$N(^2S^\prime_S)_{\frac 12^+}$&&$-\frac1{3\sqrt2}$&&*&&$-\frac 5{9\sqrt2}$&&*&&*\\[1ex]    
$\Delta(^4S^\prime_S)_{\frac 32^+}$&&0&&*&&$\frac 29$&&*&&$\frac 2{3\sqrt3}$\\[1ex]    
$\Delta(^4S_S)_{\frac 32^+}$&&0&&*&&$\frac 29$&&*&&
$\frac 2{3\sqrt3}$\\[1ex]    
$N(^2S_M)_{\frac 12^+}$&&$-\frac 13$&&*&&$-\frac 29$&&*&&*\\[1ex]    
$N(^4S_M)_{\frac 32^+}$&&0&&*&&$\frac 1{18}$&&*&&
$\frac 1{6\sqrt{3}}$\\[1ex]    
$\Delta(^2S_M)_{\frac 12^+}$&&$\frac 13$&&*&&$-\frac 19
$&&*&&*\\[1ex]
$N(^2D_M)_{\frac 32^+}$&&$\frac 13\sqrt{\frac 25}$&&
$-\frac 13\sqrt{\frac 35}$&&$\frac 29\sqrt{\frac 25}$&&
$\frac 1{3\sqrt 5}$&&*\\[1ex]    
$N(^2D_M)_{\frac 52^+}$&&$-\frac 13\sqrt{\frac 35}$&&
$-\frac 13\sqrt{\frac 25}$&&$-\frac 29\sqrt{\frac 35}$&&
$-\frac 2{3\sqrt 5}$&&*\\[1ex]    
$N(^4D_M)_{\frac 12^+}$&&0&&0&&$\frac 1{18\sqrt{5}}$&&*&&*\\[1ex]    
$N(^4D_M)_{\frac 32^+}$&&0&&0&&$-\frac 1{18\sqrt{5}}$&&0&&$\frac1{6\sqrt{15}}$\\[1ex]    
$N(^4D_M)_{\frac 52^+}$&&0&&0&&$-\frac 1{18}\sqrt{\frac 3{35}}$
&&0&&$-\frac 16\sqrt{\frac 6{35}}$\\[1ex]
$N(^4D_M)_{\frac 72^+}$&&0&&0&&$\frac 1{3\sqrt{70}}$&&0&&$\frac 1{3\sqrt{42}}$\\[1ex]
$\Delta(^2D_M)_{\frac 32^+}$&&$-\frac 13\sqrt{\frac 25}$&&
$\frac 13\sqrt{\frac 35}$&&$\frac 19\sqrt{\frac 25}$&&
$-\frac 1{3\sqrt 5}$&&*\\[1ex]
$\Delta(^2D_M)_{\frac 52^+}$&&$\frac 13\sqrt{\frac 35}$&&
$\frac 13\sqrt{\frac 25}$&&$-\frac 19\sqrt{\frac 35}$&&
$\frac 2{3\sqrt 5}$&&*\\[1ex]\hline    
\end{tabular}
\end{center}
\end{table}

\begin{table}
\caption{ The angular momentum and flavor parts of the 
helicity amplitudes for $\gamma p\to \rho^+ n$ in the
$SU(6)\otimes O(3)$ symmetry limit, while those for
$\gamma n\to\rho^- p$ are given by $A(\gamma n\to \rho^- p)=(-1)^{I+1/2}
A(\gamma p\to \rho^+ n)$, where $I$ is the isospin of the resonances.}
\protect\label{tab:(6)}
\begin{center}
\begin{tabular}{lcccccccccc}
\hline
States & & $S_{1/2}$ & & &$A_{1/2}$ & & & & $A_{3/2}$  
&\\[1ex] \hline     
 & & S & & A & & B & & A & & B\\[1ex]\hline    
$N(^2P_M)_{\frac 12^-}$&&$-\frac 2{3\sqrt{6}}$&&$\frac 
2{3\sqrt3}$&&$-\frac 4{9\sqrt{6}}$&&*&&*\\[1ex]    
$ N(^2P_M)_{\frac 32^-}$&&$\frac 2{3\sqrt{3}}$&&$
\frac 2{3\sqrt{6}}$&&$\frac 4{9\sqrt{3}}$&&$
\frac 2{3\sqrt 2}$&&*\\[1ex]    
$ N(^4P_M)_{\frac 12^-}$&&0&&0&&$\frac 1{9\sqrt{6}}
$&&*&&*\\[1ex]    
$ N(^4P_M)_{\frac 32^-}$&&0&&0&&$-\frac 1{9\sqrt{30}}$    
&&0&&$-\frac 1{3\sqrt{10}}$\\[1ex]    
$ N(^4P_M)_{\frac 52^-}$&&0&&0&&$-\frac 1{3\sqrt{30}}
$&&0&&$-\frac 1{3\sqrt{15}}$\\[1ex]    
$\Delta(^2P_M)_{\frac 12^-}$&&$-\frac 1{3\sqrt6}$&&
$\frac 1{3\sqrt3}$&&$\frac 1{9\sqrt6}$&&*&&*\\[1ex]    
$\Delta(^2P_M)_{\frac 32^-}$&&$\frac 1{3\sqrt3}$&&
$\frac 1{3\sqrt6}$&&$-\frac 1{9\sqrt3}$&&$\frac 1{3\sqrt2}$&&*\\[1ex]    
$N(^2D_S)_{\frac 32^+}$&&$-\frac 13\sqrt{\frac 25}$&&
$\frac 13\sqrt{\frac 35}$&&$-\frac 59\sqrt{\frac 25}$&&$-\frac1{3\sqrt{5}}$&&*\\[1ex]
$N(^2D_S)_{\frac 52^+}$&&$\frac 13\sqrt{\frac 35}$&&
$\frac 13\sqrt{\frac 25}$&&$\frac 59\sqrt{\frac 35}$&&$\frac2{3\sqrt{5}}$&&*\\[1ex]    
$\Delta(^4D_S)_{\frac 12^+}$&&0&&0&&$\frac 2{9\sqrt{10}}
$&&*&&*\\[1ex]
$\Delta(^4D_S)_{\frac 32^+}$&&0&&0&&$-\frac 2{9\sqrt{10}}
$&&0&&$\frac 13\sqrt{\frac 2{15}}$\\[1ex]
$\Delta(^4D_S)_{\frac 52^+}$&&0&&0&&$-\frac 19
\sqrt{\frac 6{35}}
$&&0&&$-\frac 23\sqrt{\frac 3{35}}$\\[1ex]    
$\Delta(^4D_S)_{\frac 72^+}$&&0&&0&&$\frac 2{3\sqrt{35}}$&&0&&
$\frac 2{3\sqrt{21}}$\\[1ex]    
$N(^2S^\prime_S)_{\frac 12^+}$&&$\frac 13$&&*&&$\frac 59$&&*&&*\\[1ex]    
$\Delta(^4S^\prime_S)_{\frac 32^+}$&&0&&*&&$\frac 2{9\sqrt2}$&&*&&
$\frac 13\sqrt{\frac 23}$\\[1ex]
$\Delta(^4S_S)_{\frac 32^+}$&&0&&*&&$\frac 2{9\sqrt2}$&&*&&$
\frac 13\sqrt{\frac 23}$\\[1ex]    
$N(^2S_M)_{\frac 12^+}$&&$\frac 2{3\sqrt2}$&&*&&$\frac 4{9\sqrt2}$
&&*&&*\\[1ex]    $N(^4S_M)_{\frac 32^+}$&&0&&*&&$-\frac 1{9\sqrt2}$
&&*&&$-\frac 1{3\sqrt{6}}$\\[1ex]
$\Delta(^2S_M)_{\frac 12^+}$&&$\frac 1{3\sqrt2}$&&*
&&$-\frac 1{9\sqrt2}$&&*&&*\\[1ex]
$N(^2D_M)_{\frac 32^+}$&&$-\frac 23\sqrt{\frac 15}$&&$\sqrt{\frac 2{15}}$&&$-\frac 4{9\sqrt5}$&&$-\frac 13\sqrt{\frac 25}$&&*\\[1ex]    
$N(^2D_M)_{\frac 52^+}$&&$\sqrt{\frac 2{15}}$&&$\frac 2{3\sqrt5}$
&&$\frac 49\sqrt{\frac 3{10}}$&&$\frac 23\sqrt{\frac 25}$&&*\\[1ex]    
$N(^4D_M)_{\frac 12^+}$&&0&&0&&$-\frac 1{9\sqrt{10}}$&&*&&*\\[1ex]    
$N(^4D_M)_{\frac 32^+}$&&0&&0&&$\frac 1{9\sqrt{10}}$&&0&&$-\frac1{3\sqrt{30}}$\\[1ex]    
$N(^4D_M)_{\frac 52^+}$&&0&&0&&$\frac 1{9}\sqrt{\frac 3{70}}$&&0&&$
\frac 13\sqrt{\frac 3{35}}$\\[1ex]
$N(^4D_M)_{\frac 72^+}$&&0&&0&&$-\frac 1{3\sqrt{35}}$&&0&&$-\frac 1{3\sqrt{21}}$\\[1ex]
$\Delta(^2D_M)_{\frac 32^+}$&&$-\frac 1{3\sqrt5}$&&$
\frac 1{\sqrt{30}}$&&$\frac 1{9\sqrt5}$&&$-\frac 1{3\sqrt{10}}$&&*\\[1ex]
$\Delta(^2D_M)_{\frac 52^+}$&&$\frac 1{\sqrt{30}}$&&$
\frac 1{3\sqrt5}$&&$-\frac 19\sqrt{\frac 3{10}}$&&$\frac 13
\sqrt{\frac 25}$&&*\\[1ex]\hline    
\end{tabular}
\end{center}
\end{table}

\begin{table}
\caption{ The spatial integrals in the harmonic oscillator basis.}
\protect\label{tab:(7)}
\begin{center}
\begin{tabular}{lcl}    
\hline    
Multiplet& & Expression\\[1ex] \hline   
$[70, 1^-]_1$ & &$A=\frac{3a}{2m_q\sqrt{3}}\alpha 
exp(-\frac{{\bf q}^2}{6\alpha^2})$ \\[1ex]
 & &$B=\frac{ b^\prime}{m_q}\sqrt{\frac 32}\frac{{\bf q}^2}{\alpha} 
exp(-\frac{{\bf q}^2}{6\alpha^2})$ \\[1ex]
 & &$S=\frac{\sqrt3 \mu a}{\alpha} 
exp(-\frac{{\bf q}^2}{6\alpha^2})$ \\[1ex]

$[56, 2^+]_2$ & &$A=-\frac{a}{2\sqrt{2}m_q}|{\bf q}| 
exp(-\frac{{\bf q}^2}{6\alpha^2})$ \\[1ex]
 & &$B=-\frac{ b^\prime}{2\sqrt{3}m_q}|{\bf q}|(\frac{\bf q}{\alpha})^2 
exp(-\frac{{\bf q}^2}{6\alpha^2})$ \\[1ex]
 & &$S=-\frac{ \mu a}{\sqrt6 |{\bf q}|}(\frac{\bf q}{\alpha})^2 
exp(-\frac{{\bf q}^2}{6\alpha^2})$ \\[1ex]

$[56, 0^+]_2$& &$B=\frac{ b^\prime}{2\sqrt 3m_q}|{\bf q}|
(\frac{\bf q}{\alpha})^2 exp(-\frac{{\bf q}^2}{6\alpha^2})$ \\[1ex]
 & &$S=\frac{ \mu a}{\sqrt6 |{\bf q}|}(\frac{\bf q}{\alpha})^2 
exp(-\frac{{\bf q}^2}{6\alpha^2})$ \\[1ex]

$[56, 0^+]_0$& &$B=\frac{3 b^\prime}{\sqrt2 m_q}|{\bf q}|
exp(-\frac{{\bf q}^2}{6\alpha^2})$ \\[1ex]
& &$S=\frac{3 \mu a}{|{\bf q}|}exp(-\frac{{\bf q}^2}{6\alpha^2})$ \\[1ex]

$[70, 0^+]_2$& &$B=-\frac{ b^\prime}{2\sqrt6 m_q}|{\bf q}|
(\frac{\bf q}{\alpha})^2 exp(-\frac{{\bf q}^2}{6\alpha^2})$ \\[1ex]
& &$S=-\frac{\mu a}{2\sqrt3|{\bf q}|}(\frac{\bf q}{\alpha})^2 
exp(-\frac{{\bf q}^2}{6\alpha^2})$ \\[1ex]

$[70, 2^+]_2$& &$A=\frac{a}{2\sqrt2 m_q}|{\bf q}|
exp(-\frac{{\bf q}^2}{6\alpha^2})$\\[1ex]
& &$B=\frac{ b^\prime}{2\sqrt{3}m_q}|{\bf q}|
(\frac{\bf q}{\alpha})^2 exp(-\frac{{\bf q}^2}{6\alpha^2})$ \\[1ex]
& &$S=\frac{\mu a}{\sqrt{6}|{\bf q}|} (\frac{\bf q}{\alpha})^2 
exp(-\frac{{\bf q}^2}{6\alpha^2})$ \\[1ex]
\hline    
\end{tabular}
\end{center}
\end{table}

\newpage
\subsection*{Figure Caption}
\begin{itemize}
\item Fig.1. The differential cross sections (solid curves) 
for $\gamma p\to\omega p$
at $E_\gamma=$1.225, 1.45, 1.675 
and 1.915 GeV. The data come from Ref.~\cite{saphir}. 
The $\pi^0$ exchanges are shown by the dashed curves 
and the contributions from s- and u-channel 
exclusively are shown by the dotted curves.
In (c), the dot-dashed curve represents the 
differential cross section without 
contributions from the resonance $F_{15}(2000)$.

\item Fig.2. The four single-spin polarization observables in $\gamma p\to\omega p$
are given by the solid curves at $E_\gamma=1.7$GeV. The dotted curves 
correspond to the asymmetries without the resonance $F_{15}(2000)$, 
while the dashed curves to those without the resonance $P_{13}(1900)$.

\item Fig.3. The total cross section of $\gamma p\to\omega p$ are fitted by the
solid curve with $\pi^0$ exchange taken into account. The dotted curve
describes the pure contributions from s- and u-channel. The data are
from Ref.~\cite{saphir}(triangle), \cite{ABHMC} and other 
experiments~\cite{olddata}(square).

\item Fig.4. The differential cross sections (solid curves) 
for $\gamma p\to\rho^0 p$
at $E_\gamma=$1.225, 1.305, 1.400, 1.545, 1.730 
and 1.925 GeV. The data come from Ref.~\cite{saphir}. 
The $\sigma$ exchanges are shown by the dashed curves 
and the contributions from s- and u-channel 
exclusively are shown by the dotted curves.

\item Fig.5. The four single-spin polarization observables 
in $\gamma p\to\omega p$ 
are given by the solid curves at $E_\gamma=1.7$GeV. The dotted curves 
correspond to the asymmetries without the resonance $F_{15}(2000)$.

\item Fig.6. The total cross section of $\gamma p\to\rho^0 p$ are fitted 
by the solid curve with $\sigma$ exchange taken into account. 
The dotted curve describes the pure contributions 
from s- and u-channel. The data are
from Ref.~\cite{saphir}(triangle), \cite{ABHMC} and other 
experiments~\cite{olddata}(square).

\item Fig.7. The differential cross sections  
for (a): $\gamma n\to\rho^- p$, and (b): $\gamma p\to \rho^+ n$
at $E_\gamma=$1.85 GeV. 
The data in (a) come from Ref.~\cite{benz}. 

\item Fig.8. The total cross section for (a): $\gamma n\to \rho^- p$, 
and (b): $\gamma p\to \rho^+ n$. 
The dotted line in (a) represents the cross
section for $|t|\le 1.1$ GeV$^2$. 
The data in (c) were taken with
the restriction  $|t|\le 1.1$ GeV$^2$ given by Ref.\cite{benz}.

\item Fig.9. The target polarizations for (a): $\gamma n\to \rho^- p$, 
and (b): $\gamma p\to \rho^+ n$ at $E_{\gamma}=1.7$GeV. 
The dotted lines show the results without the contribution
from the $F_{15}(2000)$.

\item Fig.10. The non-diffractive contributions from the effective
Lagrangian to the differential cross section of the $\phi$ photoproduction
at $E_\gamma=2.0$ GeV. The data come from Ref.~\cite{phidata}.
The solid and dashed curve represent the results with parameters of 
Eq.(\ref{phi1}) and (\ref{phi2}), respectively.
\end{itemize}

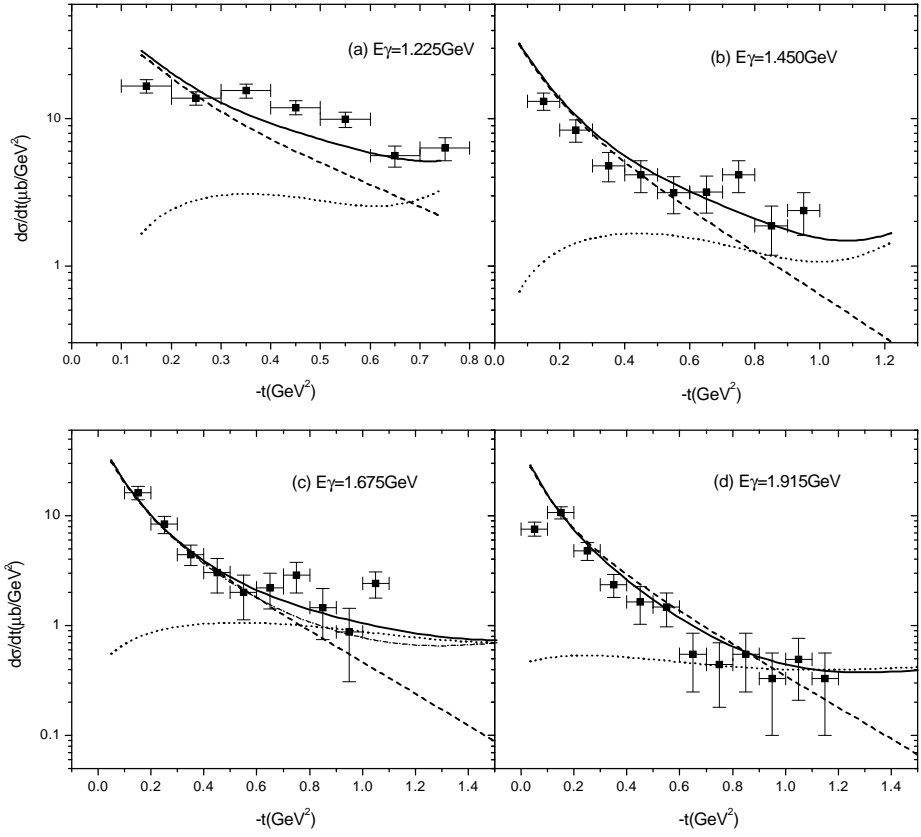
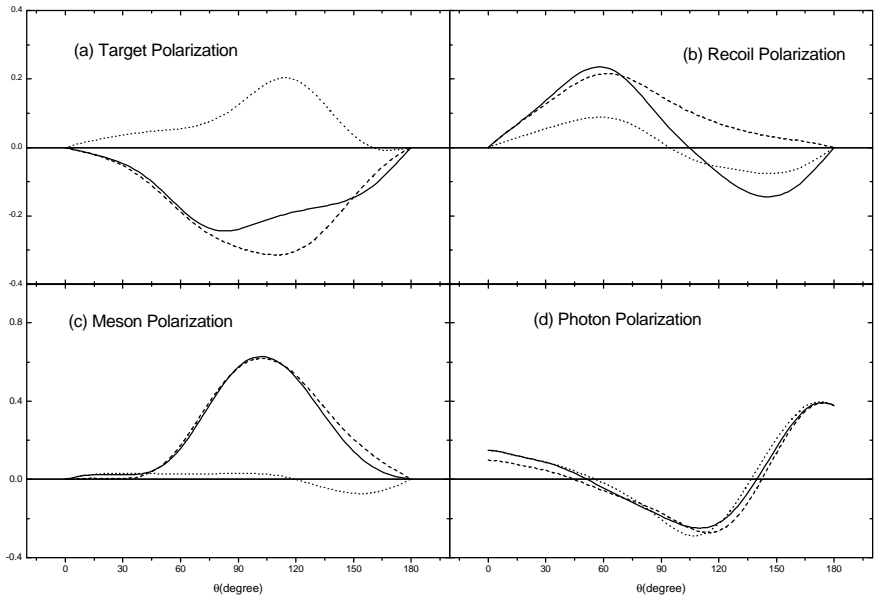
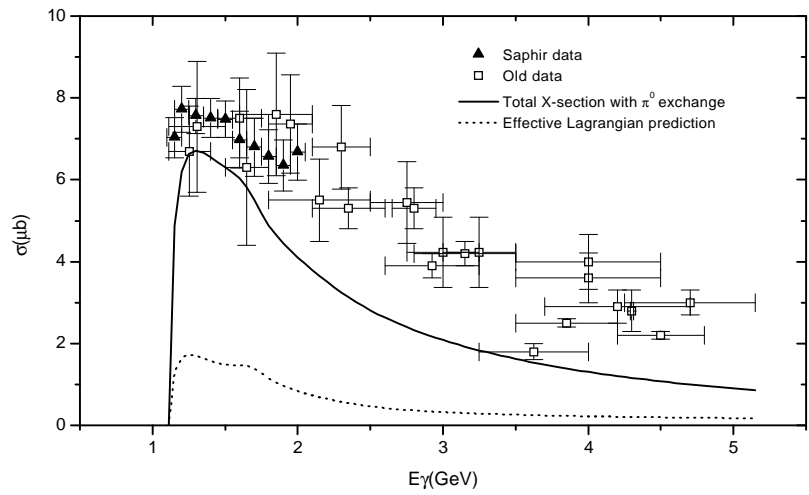
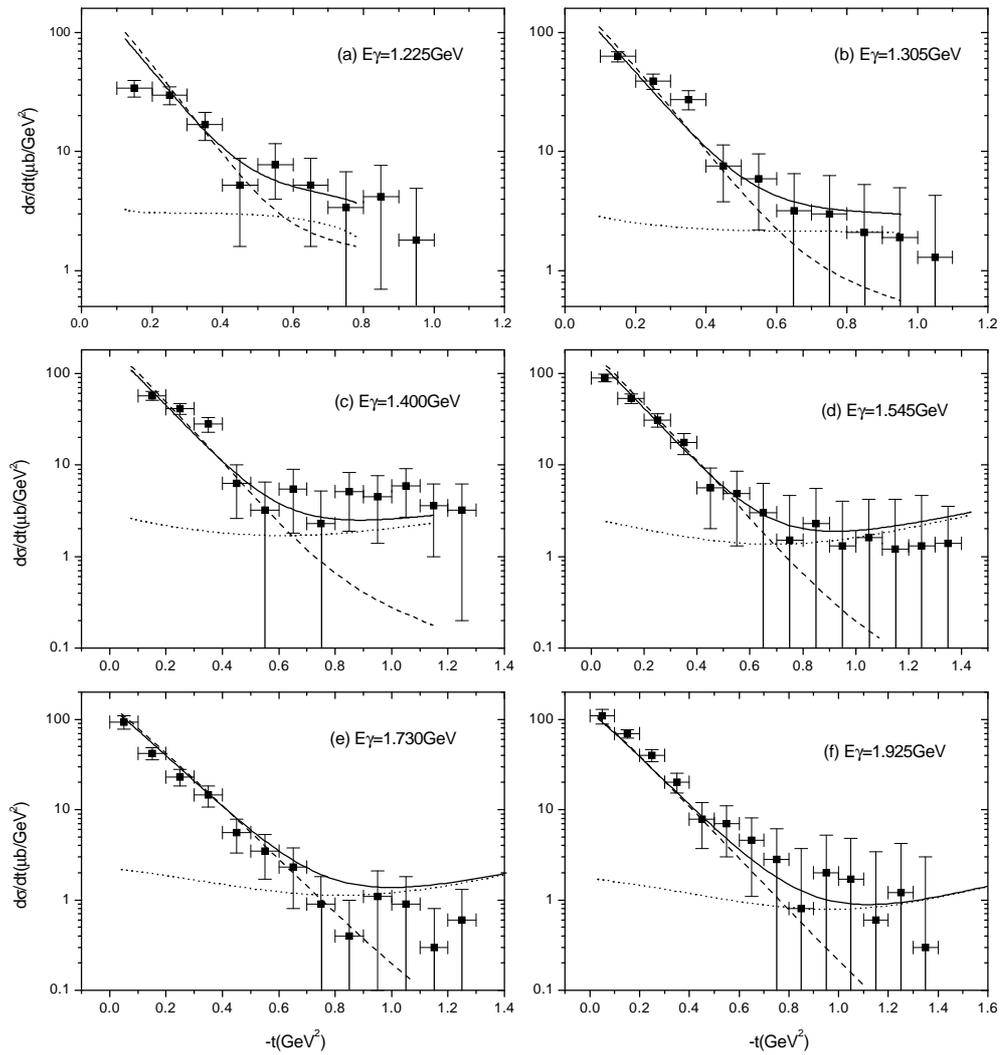
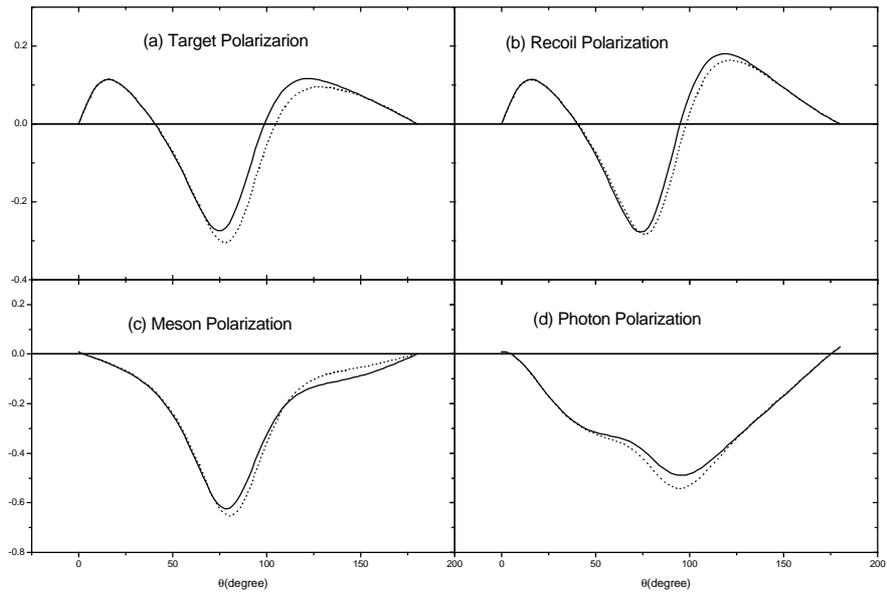
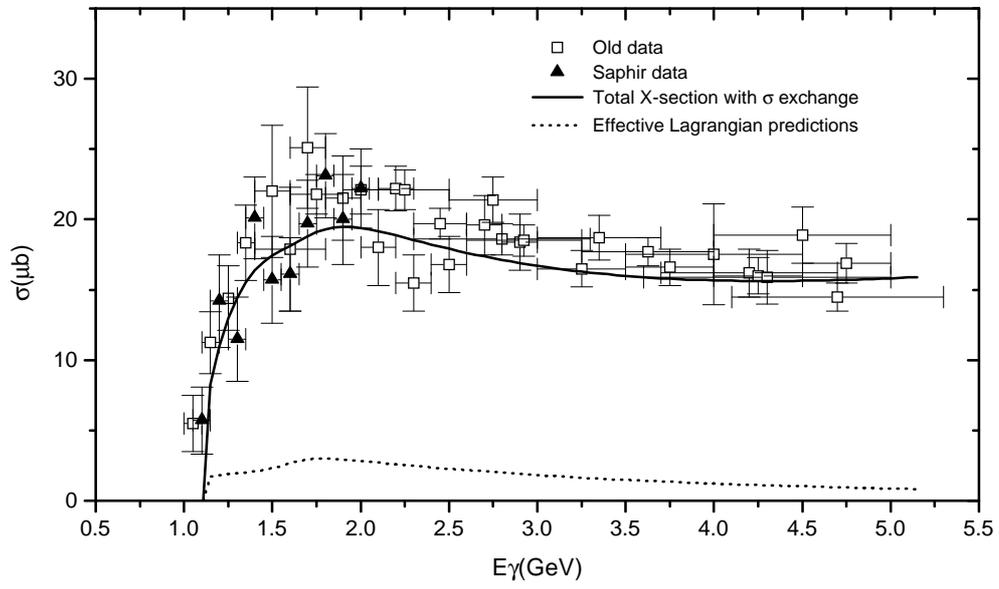
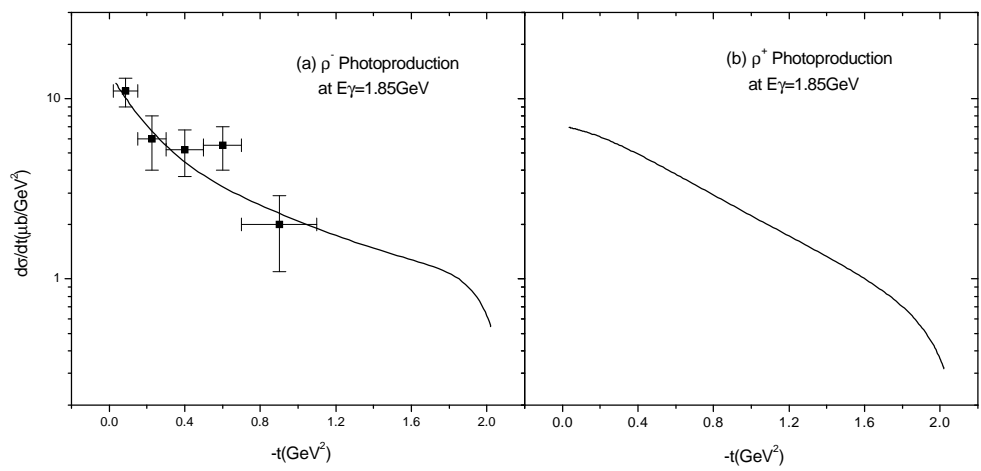
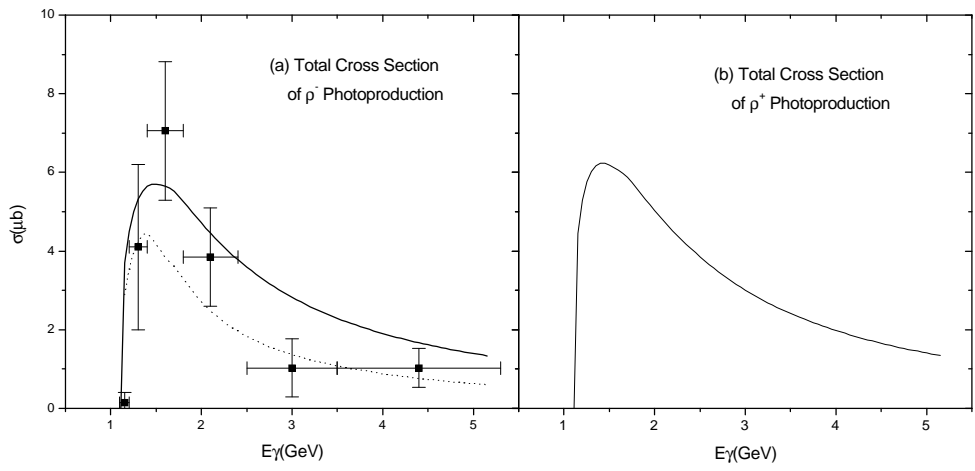
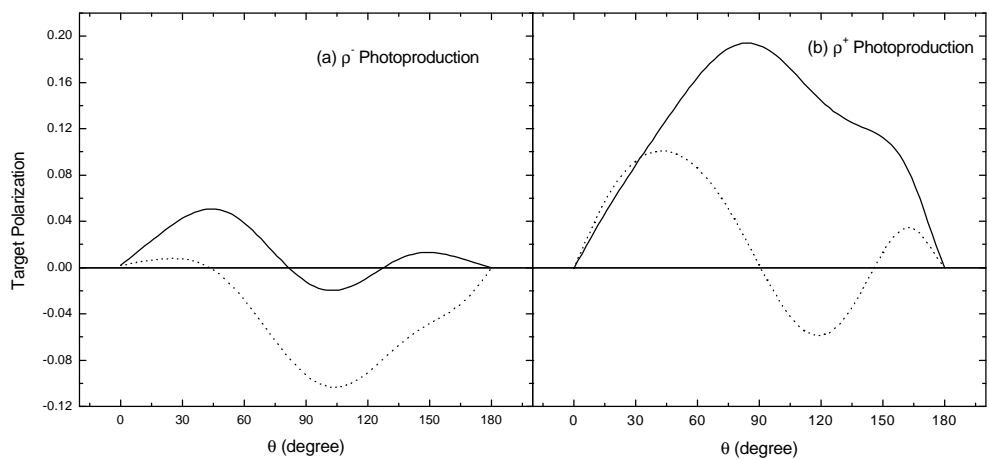
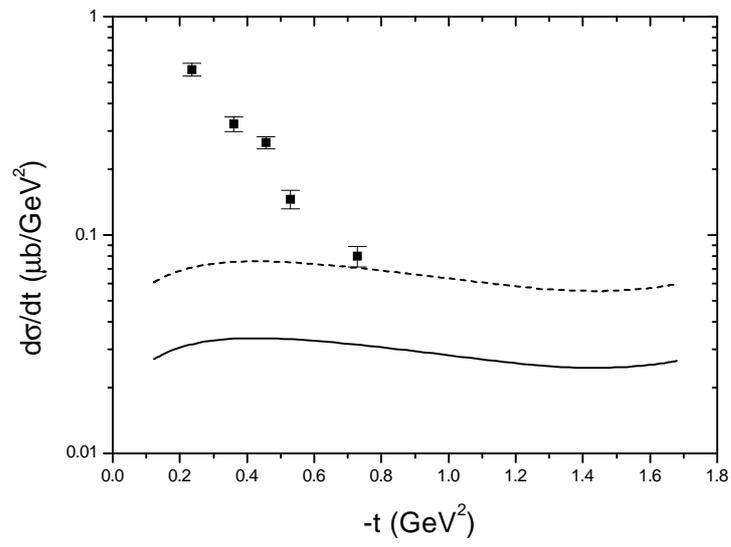
\end{document}